\documentclass[pdflatex,sn-mathphys-num,letterpaper]{sn-jnl}

\usepackage{graphicx}%
\usepackage{multirow}%
\usepackage{gensymb}
\usepackage{amsmath,amssymb,amsfonts}%
\usepackage{amsthm}%
\usepackage{mathrsfs}%
\usepackage[title]{appendix}%
\usepackage{xcolor}%
\usepackage{textcomp}%
\usepackage{manyfoot}%
\usepackage{booktabs}%
\usepackage{algorithm}%
\usepackage{algorithmicx}%
\usepackage{algpseudocode}%
\usepackage{listings}%

\theoremstyle{thmstyleone}%
\theoremstyle{thmstyletwo}%

\theoremstyle{thmstylethree}%

\begin{document}

\title{A Bridge Between Climate Science and Economics: OPTiMEM and the Heat Conjecture for Estimation of Social Cost of Greenhouse Gases}

\author*[1]{\fnm{Brian} \sur{Hanley}}\email{brian.hanley@bf-sci.com}
\author[2]{\fnm{Pieter} \sur{Tans}}
\author[3]{\fnm{Edward A.G.} \sur{Schuur}}
\author[4]{\fnm{Geoffrey} \sur{Gardiner}} 
\author[5]{\fnm{Steve} \sur{Keen}} 
\author[6]{\fnm{Adam} \sur{Smith}}

\affil*[1]{\orgname{Butterfly Sciences}, \city{Post Falls}, \state{Idaho}, \country{USA}}
\affil[2]{\orgname{Institute of Arctic and Alpine Research, University of Colorado}, \city{Boulder}, \state{Colorado}, \country{USA}}
\affil[3]{\orgname{Center for Ecosystem Science and Society, Northern Arizona University}, \city{Flagstaff}, \state{Arizona}, \country{USA}}
\affil[4]{\orgname{Chartered Governance Institute Fellow}, \city{London}, \country{UK}}
\affil[5]{\orgname{Institute for Strategy, Resilience \& Security, University College London}, \country{UK}} 
\affil[6]{\orgname{Climate Central}, \city{Princeton}, \state{New Jersey}, \country{USA}}


\abstract{
Abstract: We present an entirely new physics founded approach to estimating the social cost of carbon (SCC), the Ocean-heat-content Physics and Time Macro Economic Model (OPTiMEM) with 18 scenarios for uncertainty range. The heat conjecture assumes that weather damages curves are stochastically proportional to ocean heat increase. We model carbon combustion tightly fitted to 135 years of history, validated to datasets for greenhouse gas (GHG), temperature, and ocean heat content (OHC). We show that the social cost of 4 GHGs: CO$_2$, CH$_4$, N$_2$O and halogenated hydrocarbons, cannot be single values, but must be represented by a kind of economic phase space. We provide 1:N year loss risk models (1:10, 1:100, 1:1000) that government, engineers, and actuaries can use, and a proof that the sign of risk in losses discounting must be negative. A scenario implementing the DICE family of models implicit carbon and explicit growth assumptions, shows +18$\degree$C is breached by 2210 CE, and +110$\degree$C by 2300 CE.

We propose very long-term carbon bonds at affordable interest rates to implement real-world discounting, thus bypassing the Gordian knot of the descriptivist versus prescriptivist discounting disagreement. 

Concerns are raised about having enough low-cost fossil fuel for conversion to minimal CO$_2$ maximal energy return on energy invested (EROEI) power if nations wait too long, and low EROEI power is questioned because monetary value is dependent on energy.}




\maketitle

\section{Introduction}

We present a new approach to estimating the social cost of carbon (SCC). Here we do this as social cost of greenhouse gases (SC-GHG). This is an alternative to IAMs, developed as a collaboration, because such alternative approaches are needed to help "\textit{bridge the gulf between natural scientists and economists}" \cite{lenton2013tipping}. Thus, we developed the Ocean heat content Physics and Time Macro Economic Model (OPTiMEM) system, which implements heat-conjecture based SC-GHG. \emph{OPTiMEM is not an enumerative summation} of local weather damages across the globe. It is a strict empirical correlation model laid out in our \href{https://doi.org/10.48550/arXiv.2601.06085}{companion paper (CMP)}. We show the fit to data for each stage (\href{https://doi.org/10.48550/arXiv.2601.06085}{CMP:} fig's. 31 - 36, \& 46A). However, economists familiar with standard methods of social cost modelling are likely to think along the lines of a model, similar to an IAM, such that damages will be local, and summed together to obtain an estimate for each year, in a micro-to-macro manner. 

To implement a micro-to-macro physics based model would be a very large project requiring instrumenting very high resolution global climate models (GCMs) to estimate damages. Such an effort would have similarity to the methods of CIRA and World Weather Attribution (\href{https://doi.org/10.48550/arXiv.2601.06085}{CMP:}  \S8.7.1). In this latter method that we did not use, estimating risk requires a Bayesian approach or something similar. Running a high resolution GCM requires super-computing resources, and the Bayesian approach is not feasible at this time. We do not believe that systems like MAGICC + MESMER \cite{Beusch2022FromEmissionScenariosToSpatiallyResolvedProjectionsMAGICC_and_MESMER} can perform the equivalent of Earth System global climate models capable of detailed weather simulation because the highest resolution GCMs have questionable capacity to simulate weather at local scale \cite{Bjarke2024EvaluatingLargeStormDominanceInHighResolutionGCMsAndObservationsAcrossTheWesternUSA_2024}.  Local scale weather simulation is required for accurate weather damages estimation.

We provide methods for damages estimation based on probability/risk where this is practical (\S\ref{lb:Risk_of_single_year_damages}), both
because policy makers need to be able to make judgements of how much risk they are willing to accept, and because
actuaries, engineers, and urban planners need such methods.

In addition to new estimation methods and insights into policy, we attempt to educate climate scientists and climate economists about how each views climate. These views are quite different, and need to become congruent. We recommend reading the \href{https://doi.org/10.48550/arXiv.2601.06085}{companion article (CMP:)} to fully understand our OPTiMEM system.

\subsection{OPTiMEM climate damages heat-conjecture SC-GHG model}
OPTiMEM uses a single-cell global climate model (GCM) approach focused on Ocean Heat Content (OHC) (\href{https://doi.org/10.48550/arXiv.2601.06085}{CMP:} \S6) together with NOAA Total Weather Damages (TWD) to estimate the social cost (SC) of each of four greenhouse gases (GHGs) CO$_2$, CH$_4$, N$_2$O, and halogenated hydrocarbons (Fgases) (\href{https://doi.org/10.48550/arXiv.2601.06085}{CMP:}  \S8). \emph{The specially provided version of the NOAA Billion Dollar Disasters} \cite{NOAAbillionDlr} supplied by NOAA with complete cost data for \emph{all} recorded weather damages (TWD) (\href{https://doi.org/10.48550/arXiv.2601.06085}{CMP:}  \S4), anticipated published criticism of the dataset that has been addressed \cite{Hanley2026ReviewOfNOAAsBillionDollarDisastersInlightOfritiquingRogerPielkeJrsCriticalRemarks}. This TWD dataset covers 42 years that include changes to precipitation, storm activity, sea level, wildfire, and heat events (\href{https://doi.org/10.48550/arXiv.2601.06085}{CMP:} \S4.2, \S4.6). Thus these aspects of climate change and associated variance are present in these data. The method of finding correlates in economic data that is subject to a wide variety of influences has a long history and is common practice \cite{Lee2001PopulationInequalityHomicide,Burke2009WarmingIncreasesWarAfrica,Burke2015GlobalNonlinearEffectOfTemp,Mirza2021GlobalInequalityRemote}. The NOAA Billion Dollar Disasters data and project are now hosted by Climate Central \cite{ClimateCentralbillionDlr}.

\begin{figure}[hbt!]
\centering\includegraphics[width=5.25in]{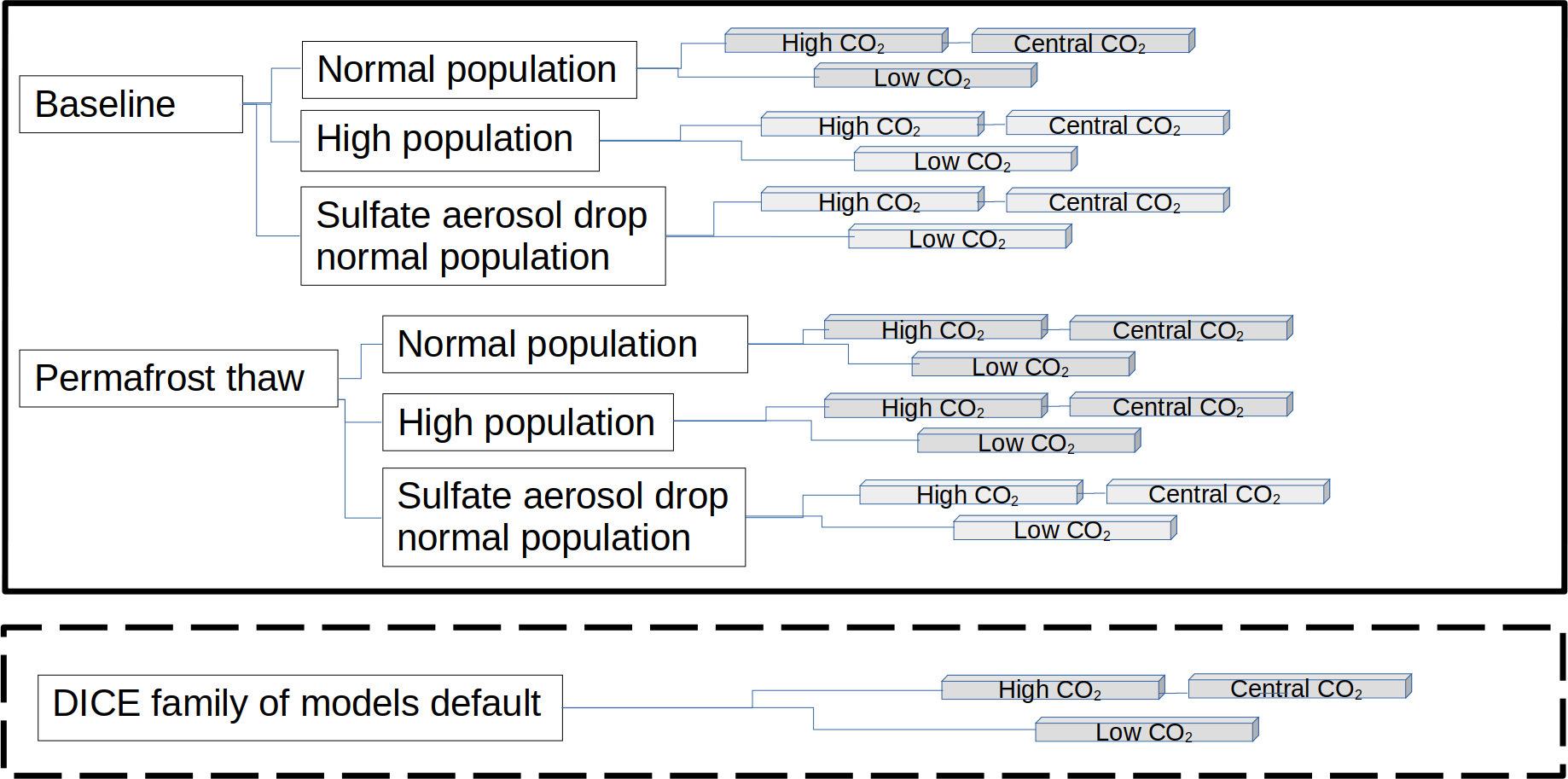}
\caption{\textbf{Scenarios tree, that sets the range of uncertainty.} Upper solid box shows the 18 primary scenarios for this study. (\href{https://doi.org/10.48550/arXiv.2601.06085}{CMP:}  \S6). Lower dashed box shows the 3 DICE family of models scenarios for the DICE implicit assumptions test. The CO$_2$ levels of high ($C_{RH}$), central ($C_{RC}$), and low ($C_{RL}$), are equations ((\href{https://doi.org/10.48550/arXiv.2601.06085}{CMP:}  \S6.2.2).}
\label{Fig_Supp_Scenario_tree}
\end{figure}

The OPTiMEM SC-GHG estimate results are 3 dimensional phase-diagrams for each year of emissions, not single numbers, because they depend upon: the uncertainty of the 18 primary scenarios (fig. \ref{Fig_Supp_Scenario_tree}), the span over which the social cost is computed, and the real dollar discount rate(s) over years, which discounting we implement using very long-term bonds in an environment where the rate of return on the bond varies (\href{https://doi.org/10.48550/arXiv.2601.06085}{CMP:}  \S3). 

We examined the literature on climate discounting, including the Stern reference \cite{Stern2007EconomicsofClimateChangeSternReview} used for DICE \cite{Nordhaus2017RevisitingTheSCC}, and find clear unsolvable disagreement (\href{https://doi.org/10.48550/arXiv.2601.06085}{CMP:}  \S2). Consequently, we created a real-world mechanism that obviates the controversy---a long-term carbon bond modelled after treasury bills (\href{https://doi.org/10.48550/arXiv.2601.06085}{CMP:}  \S3). This bond mechanism is much harder to argue with, and helps illuminate the best path forward. To account for the fluctuation of discount rates, we model a $\pm$ 2$\sigma$ range, where the standard deviation ($\sigma$) comes from 20 and 30-year T-bill history. We chose 2$\sigma$ in order to reach the DICE model's discount range of up to 5.1\% from our 1.57\% mean discount rate (\href{https://doi.org/10.48550/arXiv.2601.06085}{CMP:}  \S3.4).
 
The scenarios tree (fig. \ref{Fig_Supp_Scenario_tree}) provides the range of uncertainty, for a total of 18 primary scenarios (\href{https://doi.org/10.48550/arXiv.2601.06085}{CMP:}  \S6). This range of scenarios form the boundaries of uncertainty as shown in figure \ref{Fig_WD_Heat_Conjecture_graph}. Risk is handled by projection of statistical variance in the total NOAA weather damages dataset (fig. \ref{Fig_WD_Heat_Conjecture_Risk_graph} \& \href{https://doi.org/10.48550/arXiv.2601.06085}{CMP:} \S7.3). 
 
To accomplish this SC-GHG estimation, we developed OPTiMEM as a global climate model with a single cell based on ocean heat content (OHC). This simplified model implements our best estimate of: carbon consumption producing CO$_2$ and the drawdown of CO$_2$; CH$_4$ plus arctic permafrost release and half-life; N$_2$O and its half-life; Fgases and their half-life. Each stage: carbon $\Rightarrow$ CO$_2$ + other GHGs $\Rightarrow$ Earth Energy Imbalance (EEI) $\Rightarrow$ temperature $\Rightarrow$ OHC; is validated with empirical data (\href{https://doi.org/10.48550/arXiv.2601.06085}{CMP:}  fig.'s 23, 31, 32 \& 36).  

The OPTiMEM OHC climate model implements the leading edge of atmospheric science and (where available) the range of results from high definition global climate models, to produce the energy imbalance values that become temperature and heat content. The model itself, implemented in Maple\textsuperscript{\texttrademark} \cite{MapleCite2022-Computations}, is available. The stages of the model produce standard Excel spreadsheet files. Together, the Maple\textsuperscript{\texttrademark} standard math notation and spreadsheet output stages makes it more accessible to non-programmers, however,it is more complex than a spreadsheet. 

The heart of the heat conjecture is that the OHC curve amplitude is fitted to the TWD dataset to obtain a limited damages curve (\href{https://doi.org/10.48550/arXiv.2601.06085}{CMP:}  \S7). This fit is visible by inspection (\href{https://doi.org/10.48550/arXiv.2601.06085}{CMP:}  figs. 30, 31, 32 \& 36).  This method, in turn, allows us to compute risk (\S \ref{lb:Risk_of_single_year_damages}) (\href{https://doi.org/10.48550/arXiv.2601.06085}{CMP:}  \S7.3) based on the variance of the TWD data. This risk computation is a macro-economic statistical method based on empirical data.  

To compute damages, we scale our heat curves to achieve a binary search $R^2$ minima fit to an exponential equation fitted to our TWD dataset (fig.  \ref{Fig_WD_Heat_Conjecture_graph}) (\href{https://doi.org/10.48550/arXiv.2601.06085}{CMP:}  \S7, \S11.2). These scaled heat curves are used to project forward up to 1500 years. 

A primary task was to model reasonable limits to CO$_2$ production \cite{Tans2009_OceanicandAtmosCO2,Garrett2014LongRun1,Malanichev2018LimitsShaleOil} (\href{https://doi.org/10.48550/arXiv.2601.06085}{CMP:}  \S6.2.1, eq. 11 \& fig. 17). We use physics-based scenarios of CO$_2$ creation from realistic carbon combustion to drive a heat-equation algorithm of ocean heat content (OHC).  

The DICE scenario run with OPTiMEM implements the assumption that the economy will grow at modern historical rates, with the same energy mix as today, primarily carbon, presuming continuous carbon emissions growth driven by economic growth, for up to 300 years---we show this is impossible (\href{https://doi.org/10.48550/arXiv.2601.06085}{CMP:}  \S6.2.6 \& fig's. 16, 21).  To implement this scenario, endless carbon to support such growth was postulated, thus its carbon input is uncoupled from  any plausible real-world carbon scenario---a "Daisyworld"\cite{Watson1983Daisyworld} type of model. All of our other scenario variants (fig. \ref{Fig_Supp_Scenario_tree}) are founded on empirical realism at each step. 

We provide a feature requested by Stern \cite{Stern2021TheEconomicsofImmenseRiskUrgentActionandRadicalChange} that we have not seen in climate economics modelling, which is estimation of risk, as opposed to uncertainty. Driven by climate and weather, damages fluctuate stochastically, and smooth "averages" curves do not communicate peak event probabilities. We think individual year risk is important to provide so that economists, engineers and actuaries might have some degree of guidance going forward. To do this, we combine existing methods to generate risk curves \S\ref{lb:Risk_of_single_year_damages} (\href{https://doi.org/10.48550/arXiv.2601.06085}{CMP:}  \S7.3-7.4 fig.'s 35, 36; \S11.9-11.13 ) 

We build on what has come before, utilising basic DICE model parameters such as discount rate components and the 300 year term---amended per our analyses. We do not attempt to express non-market damages, nor other damages outside of the NOAA dataset. NOAA's dataset catalogues damage to property, insured and uninsured (\href{https://doi.org/10.48550/arXiv.2601.06085}{CMP:} \S4). 

Within each scenario, we present up to 1500 years of cost totals, with tables computed for 300, 500, and 1500 year spans (\href{https://doi.org/10.48550/arXiv.2601.06085}{CMP:}  \S13). 300 years is the DICE span. 500 years is past the curve OHC transitional phase (\href{https://doi.org/10.48550/arXiv.2601.06085}{CMP:}  \S6.5). 1500 years is our maximum. For each scenario we compute 3 different rates of CO$_2$ drawdown. And for each of those CO$_2$ drawdown rates, we use three different discount rates for our tables, which are output as .xlsx files, available as supplements, and as tables (\href{https://doi.org/10.48550/arXiv.2601.06085}{CMP:}  \S13). The full range of results is shown in isosurface graphs (fig. \ref{Fig_SCC_isosurface_2Views}). 

We make no attempt to model net-zero CO$_2$. We do not think it is politically possible for nations to  crash their economies by energy starvation. We do not see evidence that the globe will move toward net-zero carbon, except by running out of fossil fuel, which will be signalled by high prices. That decline is likely to precipitate war absent substitution. 

Our OPTiMEM heat-conjecture SC-GHG model is imperfect, as all models are in climate and economics---we can only strive for the best possible given data and methods chosen, and provide transparency to the extent possible. However, in the OPTiMEM heat-conjecture SC-GHG system we address most of the criticisms of IAMs summarised below (\S \ref{CriticalOfIAMs}). 
 
\subsection{Comparative history of the SCC}
The peak value accepted by government for calculating SCC to date is \$51 per tonne \cite[p.6]{Biden2021SocialCost_of_Carbon}. The 2050 values presented in Nordhaus' Nobel Memorial Prize lecture have a range from \$105-\$749 per tonne, presuming a 2.5$\degree$C limit \cite[Table 1 p. 2005]{Nordhaus2019TheUltimateChallengeForEconomics}. Most of the figures there have 100 and 200 year time spans. We do not think these shorter spans are necessarily valid, and their shorter term is predicated on use of discount rates that are higher than we find supportable. Thus, we make use of the more robust carbon bond mechanism, eschewing the discount debate. Let us examine as full a range of time spans and methods as can be found in literature.

The longest time span for a cost of carbon is Archer, et al's ultimate cost of carbon (UCC) which has a time scale of 10,000-200,000 years. The Archer UCC, expressed in 2020 USD, is a range of \$10,000 to \$750,000 with a mean of $\approx$ \$100,000 per tonne CO$_2$ (tCO$_2^{-1}$) \cite{Archer2020UltimateCostofCarbon}. The Archer UCC is primarily differentiated from IAM family SCC's by: A. The UCC is physics based; B. UCC uses a no-growth, zero discount economy. 

Integrated Assessment Models (IAMs) comprise DICE, iGEM, MIND, JAM, ICAM, PAGE, FUND, FAiR, and relatives. In 578 SCCs from 36 IAMs, Wang found a limited set of similar equations \cite[Table 1]{WANG2019Neta_AnalysisSCC} and SCC range of -\$13.36 tCO$_2^{-1}$ to \$2386.91 tCO$_2^{-1}$, mean of \$200.57, and $\sigma=$ $514.82$. The SCC's of the DICE family of models are characterised by having weak or nonexistent foundations in empirical data or physics. That the latter could be  true is just beginning to dawn on climate scientists. OPTiMEM is differentiated by a foundation in physics, empirical data of climate, and empirical weather damages. Our lowest range starts higher, extending above Wang for long SCC terms, and reaches or exceeds Archer's figures for time spans greater than 150-500 years and discount rates of zero and below (see fig. \ref{Fig_SCC_isosurface_2Views}). We share with Archer et al that ours is a non-IAM model estimating a price for carbon, and we also estimate a price for other GHGs.  
 
Various factors will make continuing to utilise fossil fuels at the exponentially increasing rates required to support global civilisation unlikely or impossible. OPTiMEM has a carbon consumption curve that we believe will be driven primarily by limits on supply and rising cost of extraction \cite{Tverberg2021HubbertLimit,Tverberg2021FossilFuelProblem}. (\href{https://doi.org/10.48550/arXiv.2601.06085}{CMP:}  \S6.2.1)

\subsection{Concepts, scaling and history}
Climate concepts have often been misunderstood in economics and by the public. A common founding misconception is that a global rise in average temperature is a simple scalar. This is seen in climate economics work that simply adds the average temperature increase to current weather \cite{nordhaussztorc2016Rdice, Keen2020appallingly}. The results of that economics work is collated into IPCC reports \cite[B4.6, p. 2009]{IPCC_AR6_WG3_Report}, \cite[p. 73]{Pachauri2015ipccsynthesis}. This scalar idea profoundly shapes misunderstanding of climate change impact and risk. A better conception is an exponential effect, although this too is problematic because: A. in nature exponential growth phenomena are temporary; and B. there are stochastic fluctuations stemming from the weather system and climate's foundation driver---the Earth Energy Imbalance (EEI). For instance periods of uninhabitable temperatures  \cite{Lenton2023QuantifyingHumanCostOfGlobalWarming}, or drought and flood impacts, are important stochastic fluctuations within a trend. C. A close to optimum concept of temperature is a logistic equation that has a range of positive and negative variance.

\subsubsection{Earth Energy Imbalance (EEI)}
The EEI is the net energy forcing: watts per square meter ($m^{-2}$) into or out of the planet. (The energy unit is the Joule, and the Watt is a Joule per second.) As GHGs rise, EEI rises, accelerating the rate of warming. For our purposes, EEI is the net amount of energy the earth is taking in per square meter, in one year. In the companion article are graphs showing that heat will increase until roughly 2400 CE, and beyond (\href{https://doi.org/10.48550/arXiv.2601.06085}{CMP:}  \S6.5). The rate of heat loss for 1500-2000 years afterwards depends on details of GHGs, including N$_2$O.  (\href{https://doi.org/10.48550/arXiv.2601.06085}{CMP:}  \S1.3 \& \S6).

Currently, it would require reduction of CO$_2$ equivalent GHG to 358 ppm (1993 level) to bring the EEI back into balance \cite{vonSchuckmann2020HeatstoredEarthSystem}. We are strongly opposed to the CO$_2$ equivalent concept \cite{Tans_2022UseAbuseOfC-14andC-13InAtmosphericCO2}, only using it here as an inaccurate shorthand. There is no evidence that CO$_2$ plus other GHGs should drop to this rough equivalent within the next 3000 years. Currently, no significant abatement of CO$_2$, N$_2$O, or CH$_4$ is occurring nor on the horizon, but unrealistic projections are \cite{Deprez2024CDR_SustainabilityLimitsCO2}. Abatement of halogenated gases (Fgas) is questionable.

\subsubsection{Energy of climate and weather for non-physicists}
\label{sect:HowMuchEnergy}
When GHGs collect more energy, how much energy are we talking about? A $1 \degree$C increase in global atmospheric temperature represents $\approx5 \times 10^{21}$ Joules of energy. The ocean's heat capacity is far higher. In the relatively near term the upper ocean may absorb $\approx5 \times 10^{23}$ Joules, from oceanic total capacity of $\approx 5 \times 10^{24}$ Joules per degree centigrade. The deep ocean warms slowly over time scales of thousands of years \cite{Valdes2021DeepOceanTempsThruTime, Agterhuis2022DeepSeaEoceneThermalMax2}. 

These are large numbers, but how much energy is $10^{23}$ Joules? We will look at this to clarify just how much energy GHGs collect and mostly store in the ocean. On average, a roughly similar fraction of the ocean's heat energy is expressed in weather in each year, so as that OHC energy increases, so does the absolute value of the fraction released as weather. 

A one megaton nuclear warhead is $\approx 4.18 \times 10^{15}$ Joules \cite{AtomicArchive2020}. Thus, the energy to raise the atmosphere's temperature by $1\degree$C is the energy of 1.19 million megatons of TNT (MtT). The energy to raise the upper ocean's temperature by $1\degree$C is $100$ times more, $\approx$119.6 million MtT. At an average of $\approx$0.250 MtT per warhead, and 3750 warheads, the atmospheric energy of a $1\degree$C increase is roughly 1000 times the entire US nuclear arsenal, and the upper ocean number is 100,000 times the US nuclear arsenal. On the very long time scales mentioned, the heat capacity of the deep ocean is roughly 1 million times the US arsenal for each 1$\degree$C.

Thinking about the energy involved in the earth's climate, keep in mind that in our scenarios we are talking about warming in a range from +$2.5\degree$C to +$5\degree$C (with +$15\degree$C as the presumed limit of the Paleocene---Eocene Thermal Maximum). It is heat energy which drives weather, when that heat flows across temperature gradients. Tiny fractions of total heat energy on this scale are very large. Thus we see hurricanes/typhoons, and tiny fractions can make a significant difference. 

\subsubsection{Geological climate history---empirical benchmarks from Earth's deep past}
Another primary method of considering climate effects is to compare against the background of geological climate history, because history provides a narrative yardstick for what GHG emissions can do. It is in deep history that we find analogues of extinction events correlated with climate that has similarity to our time.  

\begin{figure}[hbt!]
\centering\includegraphics[width=5.5in]{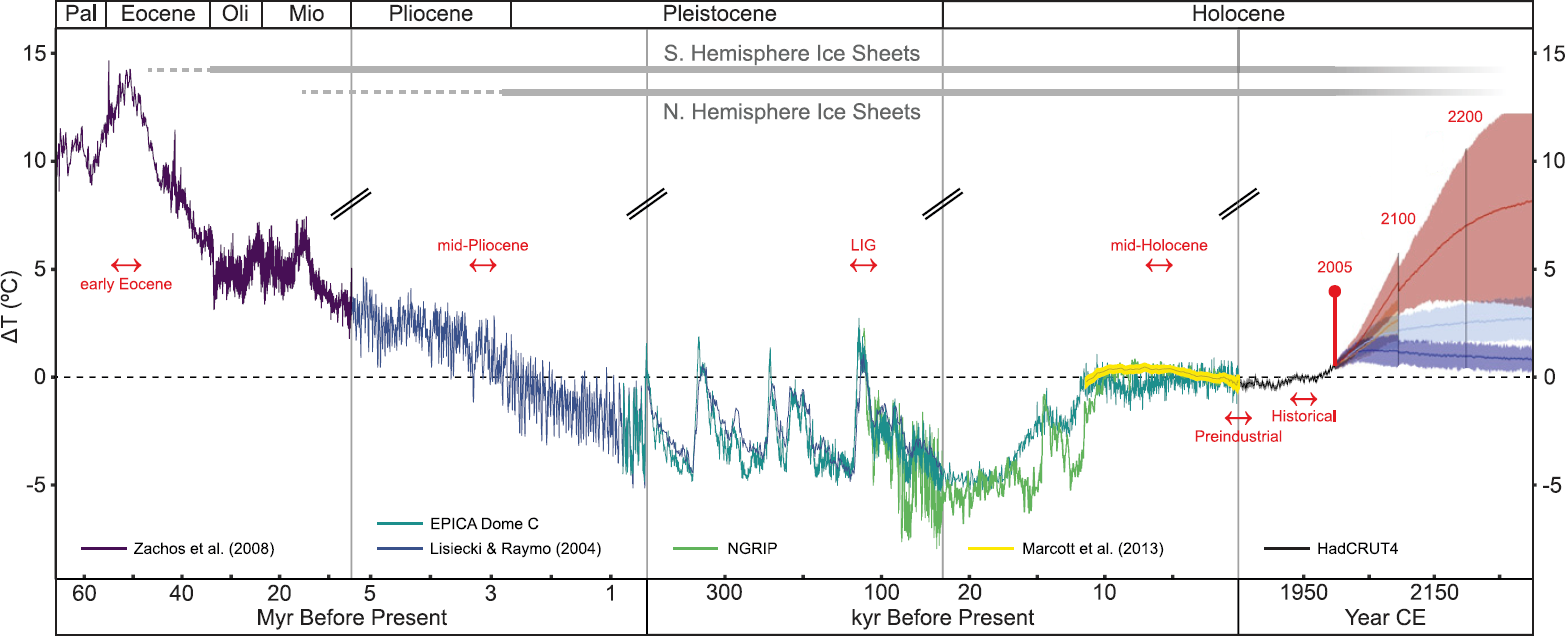}
\caption{\textbf{Fig 1 of  Burke et al  2018 \cite{Burke2018Plio} reproduced by permission in entirety with caption} Temperature trends for the past 65 Mya and potential geohistorical analogs for future climates. Six geohistorical states (red arrows) of the climate system are analyzed as potential analogs for future climates. For context, they are situated next to a multi-timescale time series of global mean annual temperatures for the last 65 Ma. Major patterns include a long-term cooling trend, periodic fluctuations driven by changes in the Earth’s orbit at periods of $10^4$ – $10^5$ y, and recent and projected warming trends. Temperature anomalies are relative to 1961–1990 global means and are composited from five proxy-based reconstructions, modern observations, and future temperature projections for four emissions pathways. Pal, Paleocene; Mio, Miocene; Oli, Oligocene. } 
\label{Fig_Burke_Eocene}
\end{figure} 

The closest periods in time are the Paleocene---Eocene Thermal Maximum (PETM) (peak temperature increase of $\approx$ +13 $\pm$ 1.5$\degree$C, 55 mya) and Pliocene (peak temperature $\approx$ +5$\degree$C, 5 mya) (fig. \ref{Fig_Burke_Eocene}). Today, the low end of Eocene CO$_2$ ($\approx$500 ppm) is $<$ 100 ppm away, and we are solidly in the central Pliocene range of CO$_2$ \cite{Burke2018Plio}\cite[Fig 5]{Rae2021Atmos} when sea level was 24 metres higher, while Eocene sea level ranged from 27-60 metres higher\cite{Hallam1984sealevel,Haq1987chronology}\footnote{Hallam and Haq diverge at 54 mya, with Haq (Vail curve) based on Exxon seismic and biostratigraphic data at 60 metres.}. However, Eocene CO$_2$ is quite insufficient to cause PETM warming. This suggests an unknown mechanism---which should  inform our uncertainty on the extreme side of damages. 

Deeper in time are two extinction events of the early Triassic. The first, following the Permian--Triassic boundary during the early Triassic at the Griesbachian--Dienerian boundary has a short time frame, on the order of 1000 years \cite{Hochuli2016}. The combined events at the end of the Permian and early Triassic caused extinction of 90\% of marine life and 75\% of terrestrial life\cite{Li2021a,Li2021b}. The Smithian period had ocean surface temperatures in the 38$\degree$ to 40$\degree$C range and an estimated rate of CO$_2$ release of 0.953 GtCO$_2$ yr$^{-1}$ \cite{Tong_Du2022103878}. Compare the Smithian period's CO$_2$ release rate with 2021's emission of 36.3 GtCO$_2$ yr$^{-1}$. In 2021, global industrial civilisation emitted CO$_2$ at 38 times the estimated Smithian-Triassic rate \cite{IEA2022CO2Emissions2021} resulting in "robust acceleration of global heating" \cite{Loeb2023MarkedIncreaseinEarthsHeatingRate,Schuckmann2023RobustEarthSystemHeating6decades}.

\subsection{Critical summary of Integrated assessment models (IAMs)}
\label{CriticalOfIAMs}

While the importance of correctly determining the social cost of carbon (SCC), is incontrovertible---despite heavy criticism of current methods comprising the DICE family of integrated assessment models (IAMs), concisely summarised below, these models remain the singular basis of influential publication in climate economics \cite{dietz2021tipping}, and the only currently accepted method for computing the SCC \cite{ExecutiveOfficeofthePresident2021ExecutiveOrder13990}. The  primary issues of IAMs are that damages functions are arbitrarily chosen with no empirical basis, calibrated to fictional data \cite{Keen2022EstimatesTippingCannotBeReconciled}, and the discount rate controversy is composed of opposed camps with discount rates from 0\% \cite[p. 10]{Dietz2008EconEthicsClimate},\cite{Stern2007EconomicsofClimateChangeSternReview,Dasgupta2021Discounting} to 5.1\% \cite{ Nordhaus2017RevisitingTheSCC, Nordhaus2018ChangesinDICE_Model,nordhaussztorc2016Rdice} (\href{https://doi.org/10.48550/arXiv.2601.06085}{CMP:}  \S2). We show that the high end discount rate has a conceptual error (\S \ref{Risk_proof}.

NOAA reported consumer price index (CPI) adjusted weather disaster damages of \$340 billion in 2017 \cite{NOAAbillionDlr}, which is 1.86\% of 2017's \$18.3 trillion GDP, at $0.8\degree$C of warming. Yet, climate economists persist into 2021 publishing estimates of climate damages in the 0.2-2\% of GDP range for up to $3\degree$C of warming. \cite{dietz2021tipping,Hsiang2017,Warren2021}. These estimates are promulgated into IPCC climate reports \cite[p. 73]{Pachauri2015ipccsynthesis}. 

Integrated Assessment Models (IAMs) comprise DICE, iGEM, MIND, JAM, ICAM, PAGE, FUND and relatives. Dietz et al 2021 \cite{dietz2021tipping} continue the tradition of ignoring virtually all criticism of IAMs, simply stating their choice not to modify 'speculative' parameters \cite[p. 44]{dietz2021supplement}, though granting that their estimates are likely lower bounds. Ignoring criticisms is also true of DICE (2016 \& 2023), as there are no meaningful changes with GAMS, which improves advancing the problematic foundations with Bayesian methods. The issues discussed below remain untouched \cite{Barrage2024PoliciesProjectionsAndTheSocialCostOfCarbon_ResultsFromTheDICE2023Model}. 

Some economists view critique of IAMs as old news in light of use of FaIR \cite{Smith_2023ClimateUncertaintyImpactsOnOptimalMitigationPathwaysAndSCC} and MAGICC \cite{Beusch2022FromEmissionScenariosToSpatiallyResolvedProjectionsMAGICC_and_MESMER}, but the climate economics field is founded upon the IAM micro-to-macro method. Despite the methods of MAGICC and FaIR, original unrevised DICE is still in use (viz. Nordhaus 2017 \cite{Nordhaus2017RevisitingTheSCC} Dietz 2021 \cite{dietz2021tipping}). And neither FaIR nor MAGICC are properly vetted from outside the rigid confines of climate economics. The 2016 \& 2023 revisions to DICE do not indicate significant departure from the 1991 survey's bias, nor anything else that has been criticised. The use of FaIR as a module used within DICE may be positive. MAGICC and MESMER may provide a better matrix. However, one has to question how a field that uses equations without basis calibrated to fiction for the past 3-4 decades is going to change that when putting damages features into MAGICC and MESMER. 

Consequently, we summarise key points of the critical literature regarding IAMs, which were created to determine impact of greenhouse gas (GHG). 

\subsubsection{Donnella Meadows}
 In 1994, Donella Meadows, in a personal letter to William Nordhaus, asked, \emph{"Why was your survey sample weighted so outrageously in favor of economists? Why was there only one ecologist, when the integrity and adaptability of ecosystems is the central link between climate and economy?"} \cite{Meadows1994LetterToWilliamNordhaus} (in ref. \cite{Nordhaus1994ExpertOpinion}).  Meadows' criticism was echoed by Keen \cite[p. 8]{Keen2020appallingly} \emph{"eight of the economists come from, ‘other subdisciplines of economics (those whose principal concerns lie outside environmental economics)’. This, ipso facto, should rule them out from taking part in this expert survey in the first place."} Meadows' letter establishes that Nordhaus was fully informed of his survey's bias in 1994. 

 \subsubsection{MERGE, the unstable foundation}
\label{sect:DICE_MERGE_WTP}
DICE uses MERGE as a foundation citation, including for the willingness to pay (WTP) value of 2\% of GDP, although MERGE is, in the words of its authors, \emph{"highly subjective"}, and contains numerous cautions and caveats, including the abstract \cite[pp. 17-18,25,28-29]{manne2005merge,manne1995merge}:
 
 Manne et al's 1995 MERGE paper is not based on empirical data. It uses estimates that the authors carefully qualify, admonishing caution, and refusing to endorse any result. (viz. \emph{"...climate impact assessment is still in a rudimentary stage. Investigators are quick to stress the preliminary nature of their findings ... The present analysis makes no attempt to narrow the range of uncertainty surrounding damages, nor do we endorse a particular set of estimates."}, continuing, \emph{"With these caveats in mind..."}\cite[p. 29]{manne1995merge}.  Manne' 1995 qualifies regarding Nordhaus \cite{nordhaus1991slow} as well, \emph{"\textbf{If} damages change quadratically with temperature..."}\cite[p. 29]{manne1995merge}. 
 
 The MERGE version cited by Dietz et al \cite{dietz2021tipping} (Manne \& Richels 2005 \cite{manne2005merge}) is essentially a manual for Manne's MERGE model\cite{manne1995merge}, stating there is no empirical foundation  \cite[p. 177]{manne2005merge}. To wit, \emph{"For market damages, we have tried to summarise the literature by supposing that a 2.5$\degree$C temperature rise would lead to GDP losses of 0.25\% in the high-income nations, and to losses of 0.50\% in the low-income nations. At higher or lower temperature levels than 2.5$\degree$C, we have made the convenient assumption that market losses would be proportional to the change in mean global temperature from its level in 2000.  For non-market damages, MERGE is based on the conjecture that expected losses would increase quadratically with the temperature rise. ...Admittedly, the parameters of this loss function are  highly speculative."} 

 \subsubsection{Stanton, Ackerman, and Kartha critique}
Stanton et al  stated that \emph{"Our review of the literature uncovered no rationale, whether empirical or theoretical, for adopting a quadratic form for the damage function – although the practice is endemic in IAMs"} \cite{STANTON2009Inside30ClimateModels}. (The quadratic damages function remains the keystone of IAMs cited in 2021, 5 years after the NAS report \cite{NAS2017Valuing-Climate-Damages-updating-SCC}".) Stanton further explains that the equations' parameters are arbitrary, equations are unjustifiedly continuous, and damages to capital stock are ignored, with damages applied only to the outputs of the productive system, resulting in unrealistic estimates of the capacity of the productive system to recover from disasters. Stanton also says that models intentionally omit equity between regions of the world, and across time, in the face of climate effects that will continue for hundreds or thousands of years. The models also use mathematically convenient, but unrealistic, oversimplifications of endogenous technological change (i.e., in the absence of government policy to promote faster change). 

\subsubsection{Weitzman’s critique}
Weitzman, 2010 \& 2012, also severely criticized the use of  quadratic functions in IAMs. Weitzman asked, \emph{"...how much we might be misled by our economic assessment of climate change when we employ a conventional quadratic damages function and/or a thin-tailed probability distribution for extreme temperatures,"}, concluding that, \emph{"the primary reason for keeping GHG levels down is to insure against high-temperature catastrophic climate risks"}\cite{WEITZMAN2010,WEITZMAN2012}. 

\subsubsection{The meticulous critiques of Pindyck}
Perhaps the most prolific and meticulous critic of IAMs, Pindyck continued in the vein of Stanton and Weitzman, with stronger language, saying IAM models are, \emph{"...so deeply flawed as to be close to useless as tools for policy analysis. Worse yet, their use suggests a level of knowledge and precision that is simply illusory"} and they are, \emph{"...completely ad hoc, with no theoretical or empirical foundation...”} \cite{Pindyck2013Whatmodelstell}. Pindyck reiterated this in 2017 \emph{"the problem goes beyond their “crucial flaws”: IAM-based analyses of climate ... can fool policymakers into thinking that the forecasts ... have some kind of scientific legitimacy"} \cite{Pindyck2017UseandMisuse}. 

\subsubsection{National Academy of Sciences report}
In 2017 the National Academy of Sciences published the results of a 31 member committee based on the criticisms above, titled "Valuing Climate Damages: Updating Estimation of the Social Cost of Carbon Dioxide"   \cite{NAS2017Valuing-Climate-Damages-updating-SCC}. Claims based on IAMs have issues with plausibility.   

\subsubsection{Lord Stern’s criticism}
Stern 2021 evaluated more "modern" IAMs \emph{"...profound implications for the policy relevance of IAMs: they provide no guidance on how to resolve differences in key judgements around risk and uncertainty, because they simply assume such differences away."} Stern also judged IAMs to be "unrobust" \emph{"...the problem is again that small changes in assumptions can yield large changes in the estimated SCC."} \cite{Stern2021TheEconomicsofImmenseRiskUrgentActionandRadicalChange}. 

\subsubsection{Keen, et al, critique}
Keen et al  pointed out that claiming 6$\degree$C of warming reduces global consumption by 1.4\% lacks any basis for credulity, and that the single calibration point at 2.5$\degree$C and 2\% damages for the quadratic criticised above is copied from Manne et al \cite{Keen2022EstimatesTippingCannotBeReconciled}. The 2\% damages comes from 1995's budget percentage for environment, which conflates actual damages to the environment with USA's 1995 willingness to pay \cite{manne1995merge}. The simple step of calibrating the quadratic through the warming level of 0.5$\degree$C in 1995 and 2\% damages for 1995 yields 100\% damages at 3.5-4$\degree$C of warming, but this result from a quadratic without basis has credibility problems as well. 

\subsubsection{The oceans boil using DICE assumptions}
DICE parameters that run to 2300 or so that show a minor drop in GDP if nothing is done are problematic.
To demonstrate the issue, we ran OPTiMEM with a classic DICE model's parameters. We set our end point as 2300,
assuming the current mix of energy and continuous economic growth driving carbon consumption, just as DICE
does. This required decoupling of OPTiMEM from limits of carbon supply. This modeling shows that using DICE
assumptions the oceans will boil (\href{https://doi.org/10.48550/arXiv.2601.06085}{CMP:}  \S6.2.6, Fig. 21). What is visible here in OPTiMEM is the benefit of using a physics founded model, because the assumptions are clearly stated, hence there is also clear policy impact, and it is possible to be specific.

\subsubsection{Summary of criticisms}
We recognise the age of the DICE family of models, also that optimum equations/algorithms may not be found on first, second, or even third attempts. We recognize that economists are not physics or chemistry trained, and do not appear to recognize when these factors are important, or how to go about representing them. Dimensional analysis is foreign to economics, which can lead to problems. Issues such as relying on fictional data for a single calibration point, arbitrary economic damages curves based on little or no empirical data, blithely stating as fact absurd results of simplistic models, and failure to improve such fundamentals when  they appear to have been known nearly 4 decades ago are more serious.

There are recent efforts to attempt improvement relative to the issues raised in this critique, albeit those attempts are built upon the same concerning foundations of the DICE family of SCC models \cite{Smith_2023ClimateUncertaintyImpactsOnOptimalMitigationPathwaysAndSCC,YANG2018225SCCUnderSharedSocioeconomicPathways,Yang2021SolelyEconomicMitigationStrategySuggestsUpwardRevisionOfNationallyDeterminedContributions}. Those foundations need to be fully examined and replaced. FaIR \cite{FaIR2025InANutshell}, or MAGICC\cite{MAGICC2025InANutshell} and MESMER \cite{Nath2022MESMER-M_anEarthSystemModelEmulator} are available to perhaps improve matters, though lacking vetting from outside the field's one-note choir. However, even in a more perfect world where tools have excellent data and algorithms, there will remain the need to find alternate approaches as a cross-check, and these should not be expected to be IAM design. That need is a primary justification for the OPTiMEM approach.
 
\section{Methods and Results}

\subsection{ Computational physics and schema }
\begin{figure}[hbt!]
\centering\includegraphics[width=5.25in]{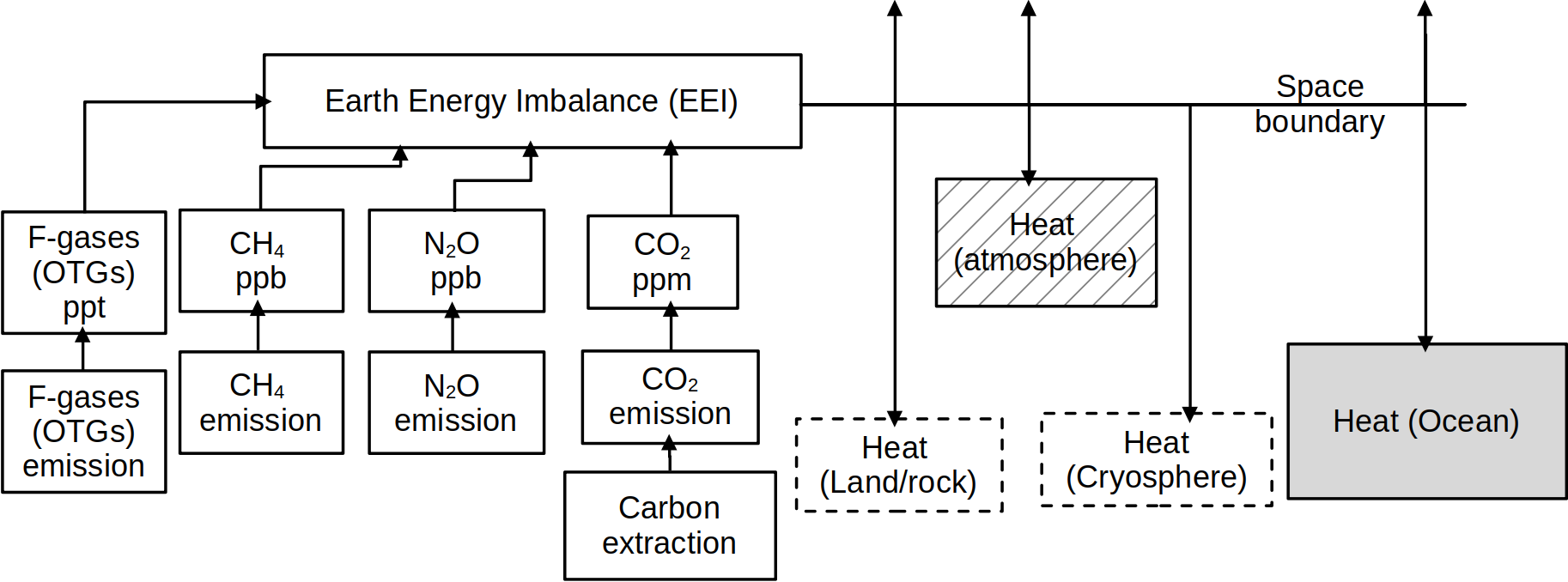}
\caption{\textbf{OPTiMEM atmospheric physics schema.} This diagram describes the climate model used for this work. Shaded ocean heat box is where heat in OPTiMEM is stored. This OHC is validated by von Shuckmann and NASA datasets, representing 88.0\% of planetary heat energy relevant to climate/weather. Cross hatched atmosphere heat box is modelled as a pseudo-surface in contact with the ocean that does not store the 0.9\% that is atmospheric heat. Dashed boxes for land and cryosphere indicate these are not used. Land and cryosphere are much more difficult to model and it was not attempted. (\href{https://doi.org/10.48550/arXiv.2601.06085}{CMP:}  \S10.2-10.3, fig. 44) We start with carbon extraction limits driving CO$_2$ emissions which translates to CO$_2$ parts per million. NOx is not a forcing of climate in OPTiMEM. N$_2$O is driven by population and Gross World Product (GWP) (\href{https://doi.org/10.48550/arXiv.2601.06085}{CMP:}  \S6.1). Methane emissions are historical fractions of CO$_2$ with the addition of permafrost emissions (\href{https://doi.org/10.48550/arXiv.2601.06085}{CMP:}  \S6.2.5). Methane converts to CO$_2$ based on its e-fold chemical lifetime (\href{https://doi.org/10.48550/arXiv.2601.06085}{CMP:}  \S6.3.2, \S11.3 \& eq. 18). Fgases also participate in changing the heat flux (EEI), which in turn heats land, ice, atmosphere, and ocean. The ocean is the largest heat reservoir by far, and the only one for which we have a usable dataset for validation. Thus, ocean heat content (OHC) is our chosen heat curve.  }
\label{Fig_Climate_Drivers_diagram}
\end{figure}

The figure \ref{Fig_Climate_Drivers_diagram} diagram  presents the schema of OPTiMEM, elaborated in \href{https://doi.org/10.48550/arXiv.2601.06085}{CMP:} \S6, and written in Maple\textsuperscript{\texttrademark} \cite{MapleCite2022-Computations}. 

\begin{figure}[hbt!]
\centering\includegraphics[width=5.5in]{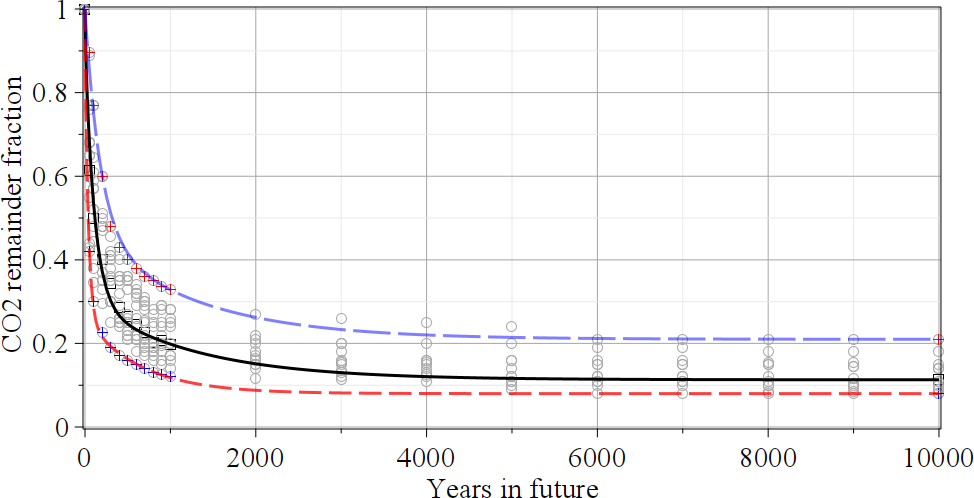}
\caption{\textbf{Remaining CO$_2$ by year fitted to Archer \cite[Fig. 1b]{Archer2009AtmosphericLifetimeOfCO2}.} Upper blue long-dash curve is fitted to points from highest Archer results C$_{RH}$.  Central solid black curve is harmonic mean of all data C$_{RC}$. Lower short-dash red curve is fitted to lowest remainder points, C$_{RL}$. Grey circles are density plot of Archer data sampled at intervals. Upper and lower crosses are calibration points for upper and lower curve fits. Central boxes are computed calibration points for harmonic mean. These three CO$_2$ remainder equations provide the fraction for future years. (\href{https://doi.org/10.48550/arXiv.2601.06085}{CMP:} \S6.2.2 eq’s. 12, 13, 14).}
\label{Fig-RemainderFractions}
\end{figure}

We derive our empirical estimates of partial economic climate damages using datasets comprising: NOAA complete damages to include all loss costs \cite{NOAAbillionDlr} (\href{https://doi.org/10.48550/arXiv.2601.06085}{CMP:}  \S4), CO$_2$  \cite{Bereiter2015RevisionOfEPICA,MaunaLoaCO2}, N$_2$O, CH$_4$, \& trace gases \cite{NOAA2022AGGIClimateTable}, ocean heat content \cite{vonSchuckmann2020GCOSEHIEXPv2}, inflation \cite{FRBM1913-2021CPI}, Federal reserve discount rate \cite{FRED2021INTDSRUSM193N}, US GDP \cite{FRED1947-2021GDP}, Gross world product (GWP) \cite{FRED2021NYGDPMKTPCDWLD}, US Federal expenditures \cite{FRED2020FGEXPND}, US defence spending \cite{FRED1947-2021FDEFX}, quantitative easing \cite{FREDqe2008},  and similar data. We use higher precision than required to minimise potential numerical error carrying through many iterations, 20 digits for our first phase, 12 digits for the second. Our estimate of CO$_2$ absorption by the ocean/biosphere/geology is nuanced, with high, harmonic mean, and low absorption based on a review of 8 climate models by Archer \cite{Archer2009AtmosphericLifetimeOfCO2} (fig. \ref{Fig-RemainderFractions} ), yielding three long-term curves (remainder curves) for CO$_2$ remaining in the atmosphere after emissions:  low, central, and high (\href{https://doi.org/10.48550/arXiv.2601.06085}{CMP:}  \S6.2.2 \& eq.'s 12-14 ). 

\label{lb:CMIP}
Central and low values for all scenarios fall within the central range SSP245 of CMIP6 temperature \cite{CMIP62023ClimateProjections,CMIP62026CopernicusDataTutorials} (\href{https://doi.org/10.48550/arXiv.2601.06085}{CMP:}  Panel B of fig's. 24-29). High range scenarios may land in the lower end of SSP585, exceeding +4$\degree$C and reaching +5$\degree$C (\href{https://doi.org/10.48550/arXiv.2601.06085}{CMP:}  Panel B of fig's. 25, 26, 28, 29). See also Burke (fig. \ref{Fig_Burke_Eocene}) for its projection of CMIP6 out to 2300 CE. Over time, CMIP projections have been a bit lower than empirical data \cite{Carvalho2022CMIP3_CMIP5_CMIP6_recent_future_climate_projections}. This suggests that our modelling is reasonable.

\label{eqs:NetCO2Remainders}

We provide methods for damages estimation based on probability/risk where this is practical (\S\ref{lb:Risk_of_single_year_damages}), both because policy makers need to be able to make judgements of how much risk they are willing to accept, and because actuaries, engineers, and urban planners need such methods.  

\subsection{The heat conjecture}
The essence of the heat conjecture (\href{https://doi.org/10.48550/arXiv.2601.06085}{CMP:}  \S7) is to use a curve founded in physics that is directly driving climate and weather events to provide a reasonable curve for damages from climate change. The logic of this is that a similar fraction of the heat in the upper ocean should be available to create weather in any given year---subject to stochastic variance. So, the magnitude of that proportion of OHC should rise, on average, synchronously with OHC. Thus, scaling the OHC to fit existing data should provide as good a measure of future projection as anything available to us. 

A temperature curve was considered, but OHC is the primary driver/moderator of weather \cite{Raghuraman2024WarmingSpikeElNinoSouthernOscillation,Watson-Parris2024SO2ConfoundedByInternalVariability,Samset2024_2023TempsReflectSSTVariability}, and should remain so. The temperature curve rises and falls faster than global heat content by centuries, without reaching system equilibrium. (\href{https://doi.org/10.48550/arXiv.2601.06085}{CMP:}  \S6.5, fig.'s 24-29; Compare $\Delta$T B panels with OHC (Q) C panel, and $\Delta$Q OHC heat flow D panel. And fig. 45.) Thus, using a temperature curve ($\Delta$T) should dramatically over-estimate the rapidity of overall climate change. More concerning, is that reliance on a $\Delta$T curve will wildly underestimate the persistence of climate. Similarly, GHGs will also rise and fall, long before equilibrium can be reached. This is because the heat capacity of the ocean is immense (\S \ref{sect:HowMuchEnergy})). It is also worth noting that if all GHGs were abated, climate damages should have a period of calm, then ramp up when the ocean began releasing immense amounts of energy every year for centuries (\href{https://doi.org/10.48550/arXiv.2601.06085}{CMP:}  fig. 15 panel D \& fig. 45). 

\begin{figure}[hbt!]
\centering\includegraphics[width=5.25in]{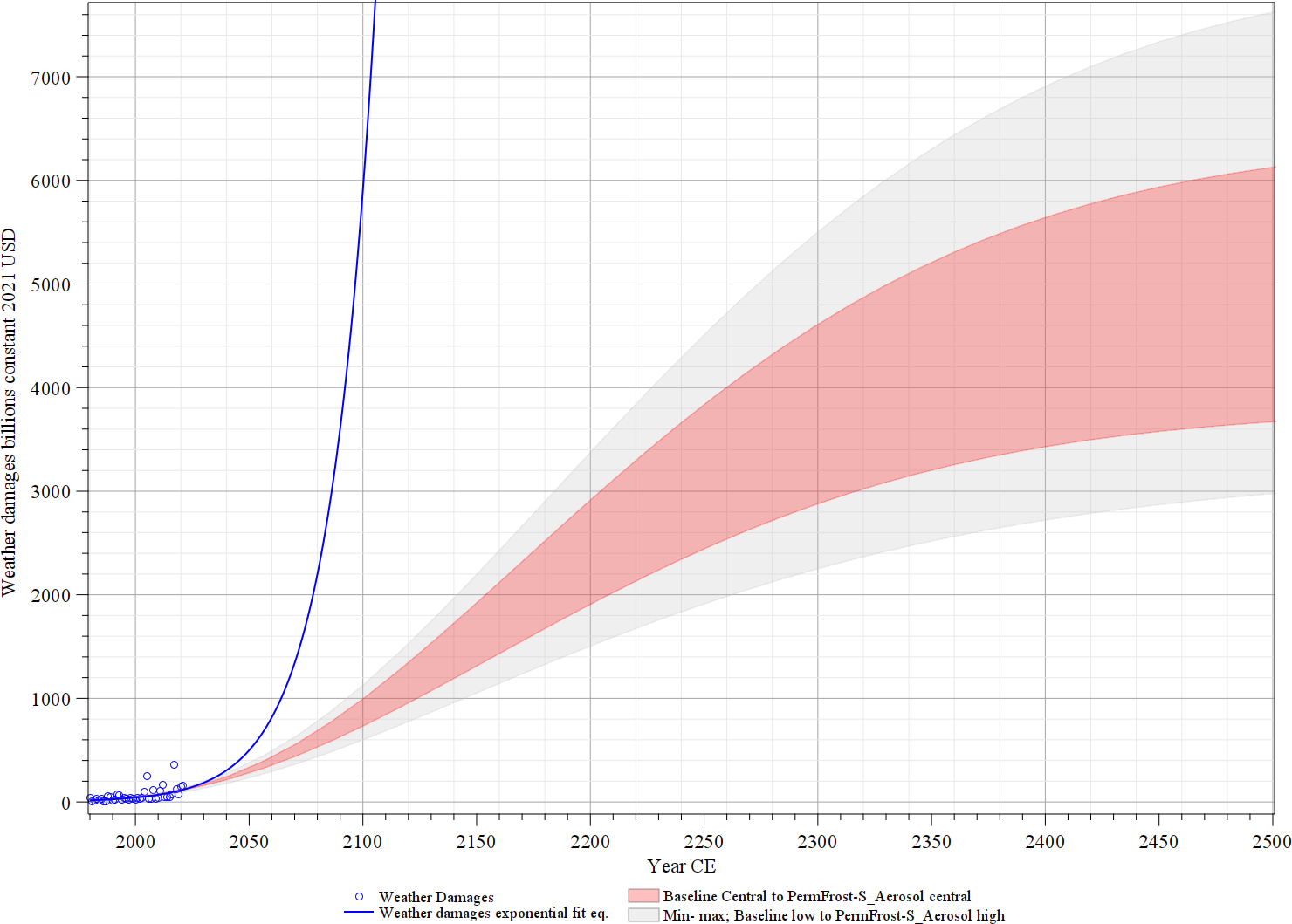}
\caption{\textbf{Heat conjecture weather damages curves estimating uncertainty. } This graph is a slice at Tab A through figure \ref{figPVDrateNumYears}, extending approximately 475 years. Blue datapoints, TWD (NOAA total weather damages). Blue curve is the trajectory of $WDe(y)$ (\href{https://doi.org/10.48550/arXiv.2601.06085}{CMP:}  \S11.11) exponential fit equation to TWD data, provided to show the fit to the scaled OHC curves. Shaded curve regions are weather damages curves. Light grey region, shows the outer limits from lowest to highest scenarios. Inner peach colored region shows the high and low central scenarios.}
\label{Fig_WD_Heat_Conjecture_graph}
\end{figure}

Thus, the range of curves for OHC results are fitted to the region of NOAA data, and scaled commensurately as shown in figure \ref{Fig_WD_Heat_Conjecture_graph}. These resulting weather damages curves provide the basis for forward projection of OPTiMEM's heat-conjecture SC-GHG model.

\subsubsection{Uncertainty regarding ocean heat content (OHC) }
The uncertainty of OHC in OPTiMEM is strictly a matter of the quantity of each greenhouse gas emitted, through the dependencies that drive GHG emissions, unless EEI conditions change. Everything else is physics. Thus, the OHC curves for different scenarios can be computed, and nothing except gas emissions are uncertain. (We exclude black swan events like massive volcanic activity and asteroid strikes.) In the companion article we describe how OPTiMEM drives emissions from best available literature's basis for the equations that describe each gas' emission (\href{https://doi.org/10.48550/arXiv.2601.06085}{CMP:} \S6). 

Carbon modelling (\href{https://doi.org/10.48550/arXiv.2601.06085}{CMP:}  \S6.2.1) presents equations fitted to historical records that make use of equation forms used to project carbon production from wells and mines. These equations drive CO$_2$ emissions. Three new equations to model the uncertainty of CO$_2$'s unique gas behaviour are defined (\href{https://doi.org/10.48550/arXiv.2601.06085}{CMP:}  \S6.2.2, \S6.3.1). This CO$_2$ section also presents modelling of CH$_4$ that includes permafrost melt (\href{https://doi.org/10.48550/arXiv.2601.06085}{CMP:}  \S6.2.5). Ice melt is an area in flux, with modelling not caught up to observations (\href{https://doi.org/10.48550/arXiv.2601.06085}{CMP:}  \S9.5).  OPTiMEM's heat-conjecture SC-GHG model improves significantly on current models for nitrogen (\href{https://doi.org/10.48550/arXiv.2601.06085}{CMP:}  \S6.1). Herein, we make use of the most recent science accounting for nitrogen \cite{Tian2020N2OSources_sinks}. 

\subsection{Risk of single year weather damages exceeding the norm} 
\label{lb:Risk_of_single_year_damages}
The total NOAA weather damages data used for the OPTiMEM heat conjecture method ($fWD_m(y)$ eq. (\href{https://doi.org/10.48550/arXiv.2601.06085}{CMP:}  \S11.2)) vary by year. We quantify risk using the deviations from fitted norm curve (\href{https://doi.org/10.48550/arXiv.2601.06085}{CMP:}  \S7.3 \& \S11.9). These risk curves are used to estimate probability of an outlier year with damages at or above the level shown by the graphs. Actuaries might use this also for odds of outlier year losses to cover. 

\begin{figure}[!ht]
\centering\includegraphics[width=4.5in]{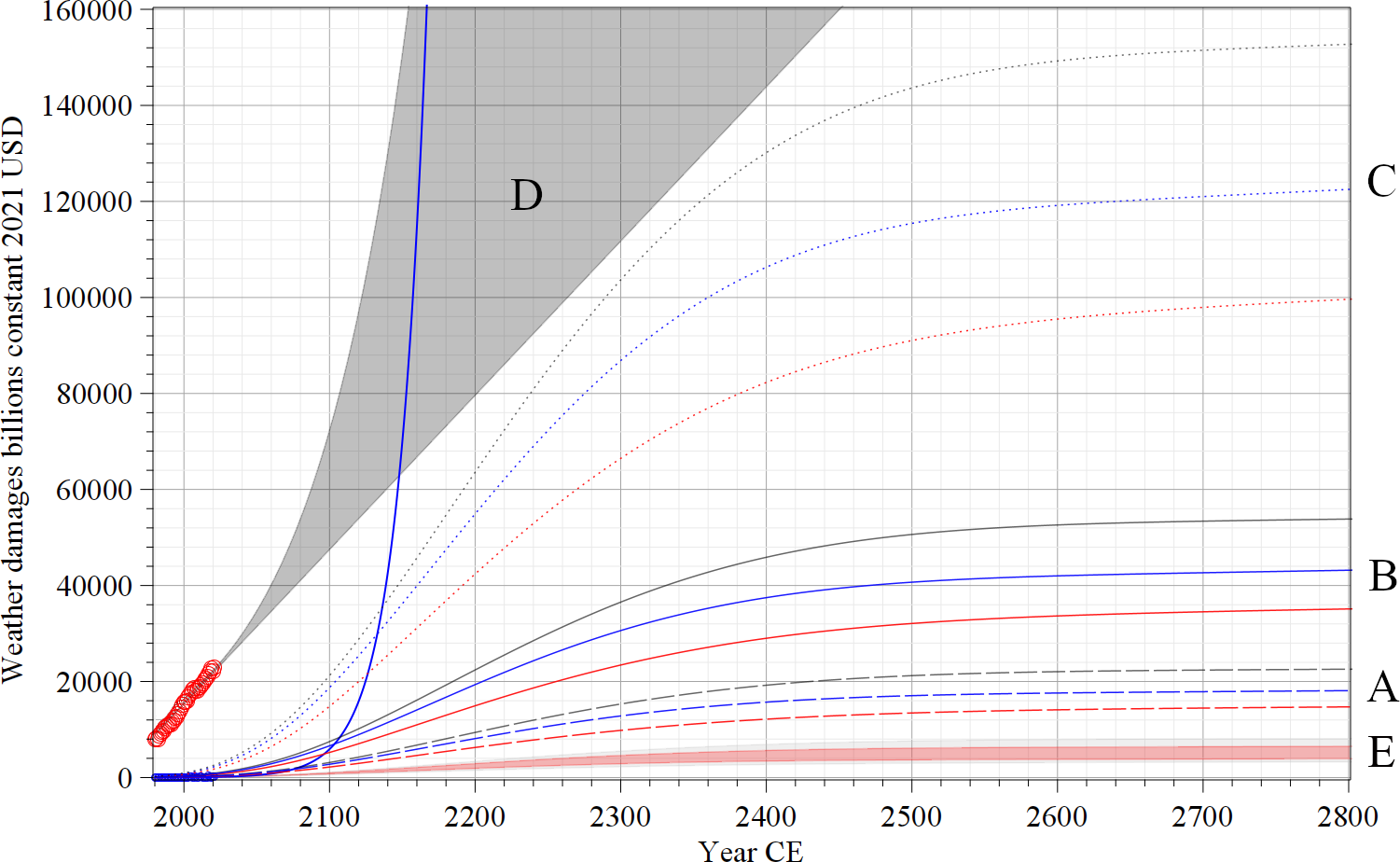}
\caption{\textbf{$\sigma$ risk curves based on the uncertainty spread of figure \ref{Fig_WD_Heat_Conjecture_graph}.} This is a slice through Tab A across figure \ref{figPVDrateNumYears} \textbf{These curves show the risk for outlier weather damages in any single year compared to projected GDP growth. These are projections from the NOAA dataset using the heat conjecture curves with $\sigma$ risk added (eq. \ref{eq:EsigmaRisk}). No discount rate is applied, and no cost summation across years is done.} Dashed curves \textbf{A} (lowest group), are 1:10 years (10\%) risk, for low, central, and high CO$_2$. Solid curves \textbf{B} (middle group) are: 1:100 years (1\%) risk, for low, central, and high CO$_2$. Dotted curves \textbf{C} (upper group) are: 1:1000 years (0.1\%) risk, for low, central, and high CO$_2$. Grey shaded region \textbf{D} shows US GDP from a low linear projection to a high 1.48\% per year growth. Shaded region \textbf{E} is the projection of average damages shown in  detail in figure \ref{Fig_WD_Heat_Conjecture_graph} (and \href{https://doi.org/10.48550/arXiv.2601.06085}{CMP:}  \S7.2-7.4). Red circles are datapoints for US GDP. Blue solid circles are NOAA weather damages, and the heavy, solid blue exponential curve is the least squares curve fit $WDe(y)$ eq. (\href{https://doi.org/10.48550/arXiv.2601.06085}{CMP:}  \S11.11). The solid blue curve is projected well past its validity region, which extends no farther than 2080 CE. Note that lower discount rate risk curves could potentially impinge or cross beyond the upper limit of region \textbf{D} (not shown). }
\label{Fig_WD_Heat_Conjecture_Risk_graph}
\end{figure}

To generate risk curves, as with base weather damages, exponential curves are fitted to positive standard deviation points. Heat curves are then fitted using the binary search $R^2$ minima algorithm over the NOAA weather damages span, generating a y axis scalar of 0.3626 with $R^2$ = 0.9563. To find risk we apply Chebyshev's theorem (\href{https://doi.org/10.48550/arXiv.2601.06085}{CMP:}  \S11.10) to obtain a multiplier, and add the resulting curve to weather damages as shown in figure \ref{Fig_WD_Heat_Conjecture_Risk_graph}. 

In figure \ref{Fig_WD_Heat_Conjecture_Risk_graph} three risk levels ($r$) are shown for individual years:  A. $r = 10\%$ 1:10 ; B. $r = 1\%$ 1:100 ; and C. $r=$ 0.1\% 1:1000. Computation of ($k \cdot \sigma$) (\href{https://doi.org/10.48550/arXiv.2601.06085}{CMP:}  \S11.10) provides the $\sigma$ multiplier for each risk level ($r$) to compute dollar risk from extreme weather in future years.  These are, respectively: \\ 
\begin{flalign}
  \begin{aligned}  
\label{eq:EsigmaRisk} 
 &(2.236 \times E\sigma(y) ) + fWD_m(y) \text{ for } r=10\% \text{ (1:10)}  \quad (7.07 \times E\sigma(y) ) + fWD_m(y) \text{ for } r=1\% \text{ (1:100)} \\
 &(22.36 \times E\sigma(y) ) + fWD_m(y) \text{ for } r=0.1\% \text{ (1:1000)} \\ 
&Where: \quad y= \text{year,} \quad r= risk \quad r\text{ term is $k$ result of Chebyshev eq. (\href{https://doi.org/10.48550/arXiv.2601.06085}{CMP:}  \S11.10)} \\  
& \qquad \quad E\sigma(y) = \text{ risk effect $\sigma$ eq. (\href{https://doi.org/10.48550/arXiv.2601.06085}{CMP:}  \S11.9)}  \\ 
& \qquad \quad fWD(y) = \text{ weather damages functions eq. (\href{https://doi.org/10.48550/arXiv.2601.06085}{CMP:}  \S11.2)} \\
\end{aligned}
\end{flalign}

 We see in figure \ref{Fig_WD_Heat_Conjecture_Risk_graph} that 0.1\% risk (1 in 1000, dotted curve set C) for a single year appears uncomfortably close to the 2025's GDP region $\approx 2200-2350$ CE. Not shown is risk for slices through figure \ref{figPVDrateNumYears} below the 0\% contour---it is obvious that these could extend to overwhelm the economy.  

\subsubsection{Risk over time spans}

Risk curve sets A, B and C of fig. \ref{Fig_WD_Heat_Conjecture_Risk_graph} show the probability in any single year that weather damages for that year will be $\geq$ to a specified curve. To convert single year risk to risk over some span of years, we use eq. \ref{eq:ChebyshevRiskOverY}, below.
 Generally, to consider risk, policy makers or actuaries want to know what the risk is within some time period such as 30, 50, or 100 years, and rarely, 1000 years. To compute this, use eq. \ref{eq:ChebyshevRiskOverY}. Results are shown in table \ref{TableRiskOfRareWeatherEventYears} as percentages to 2 decimal places, which makes clear that even 1 in 1000 weather years are non-trivial probabilities when considered over 100 year time spans.

\begin{table}[hbt!]
 \centering
 \caption{Rare weather year risk over 10 to 1500 year time spans: $R_S$.  Combine with magnitude from figure \ref{Fig_WD_Heat_Conjecture_Risk_graph}. First column is risk on a 1-year basis. Across the top to the right of risk is time span of risk computation.} 
\label{TableRiskOfRareWeatherEventYears}
\begin{tabular}{c|cccccccc}
  & \multicolumn{8}{c}{Span of years for risk computation} \\
$r$      & 10      & 30      & 50      & 100      & 300      & 500      & 1000     & 1500     \\ \hline
10.0\% & 65.13\% & 95.76\% & 99.48\% & 100.00\% & 100.00\% & 100.00\% & 100.00\% & 100.00\% \\
1.0\%  & 9.56\%  & 26.03\% & 39.50\% & 63.40\%  & 95.10\%  & 99.34\%  & 100.00\% & 100.00\% \\
0.1\%  & 1.00\%  & 2.96\%  & 4.88\%  & 9.52\%   & 25.93\%  & 39.36\%  & 63.23\%  & 77.70\%  \\ \hline
\end{tabular}
\end{table}

\begin{align}
\begin{split} \label{eq:ChebyshevRiskOverY}
 R_S &= (1-(1-r)^S)   \\
\text{Where: } R_S &= \text{the probability that one or more extreme years occur in the span of years} \\ 
r &= \text{ risk used in Chebyshev theorem and }0 \leq r \leq 1  \text{   (\href{https://doi.org/10.48550/arXiv.2601.06085}{CMP:}  \S11.9-11.10})\\
S  &= \text{ span  of  years }  \\
\end{split}
\end{align} 

\begin{quote}
 Example: $ r = 1\% \textbf{,  } S = 100$  \\
\centering $  R_S =  1-(1-0.01)^{100}$  \\
\centering $ \quad R_S = 0.6339$ \quad  or \quad 63.4\% \\
\end{quote}

\subsection{Present value for future losses }

In our computation of future losses, the time value of money discount rate is implemented using a carbon bond. The concept is that for a tonne of each GHG emitted, a bond would be purchased from the Treasury for the SC-GHG term selected by regulators, or simply allocated for spending by the government on abatement or high EROEI zero-carbon energy. We presume the minimum SC-GHG term is 300 years in practice, and show graphical results with terms from 10 years to 1500 years (fig. \ref{figPVDrateNumYears}, \ref{Fig_SCC_isosurface_2Views}, eq. \ref{eq:PVFLisosurface1}). 

\subsubsection{A proof that the sign of risk adjustment on the exponent for losses must be negative (opposite to risk adjustment for gains) }
\label{Risk_proof}
Let us begin with a project that has an \textbf{expected} positive revenue stream $E(t)$, and an \textbf{actual} positive revenue stream $A(t)$ \cite{Hanley2022SignOfRiskForFutureLosses}. At an arbitrary value of $t$ (where $t=$time, $E(t)$ is compared with $A(t)$. We will assume that both $E(t)$ and $A(t)$ start at the same initial value of income $I$, and grow exponentially at different rates $g_{_E}$ (expected growth) and $g_{_A}$ (actual growth) respectively. The expected and actual returns are:
\begin{align}
E(t)&=I \cdot e^{(g_{_E}) \cdot t}   &  A(t)&=I \cdot e^{(g_{_A}) \cdot t} 
 &\text{Where: } g_{_E} \text{ and } g_{_A} > 0
\end{align}

The discounted value of those cash flows subtracts a \emph{positive} term for the \textbf{time value of money}, $d_{tvm}$, from both exponents, to account for the lower present value of future income versus income today. Thus we have \textbf{expected} $E_{tvm}$ and \textbf{actual} $A_{tvm}$ time value of money. 
\begin{align} \label{eq:Etvm_Atvm1}
E_{tvm}(t)&=I \cdot e^{(g_{_E} - d_{tvm} ) \cdot t}   &  A_{tvm}(t)&=I \cdot e^{(g_{_A} - d_{tvm}) \cdot t} 
& \text{Where: } E_{tvm}(t) > A_{tvm}(t) > 0 
\end{align}

Let us next consider the relevant risk in the ordinary present value case, that actual returns will be below expected returns. It is possible for a project with a positive future income stream to be net present value positive on a time value of money basis, but lower---or even negative---on a risk adjusted basis. Thus, a \textbf{risk adjustment rate} $d_{rar}$, is applied to $E_{tvm}(t)$ to cover the case where returns are \emph{lower} than expected, yielding $E_{rar}$. There is no need to apply $d_{rar}$ to actual returns, because \emph{actual returns} are the basis for determining what the correct fit value of $d_{rar}$ is.
\begin{align} \label{eq:Etvm_Atvm2}
E_{rar}(t)&=I \cdot e^{(g_{_E} - d_{tvm} - d_{rar} ) \cdot t}   &  A_{rar}(t) &= A_{tvm}(t) =I \cdot e^{(g_{_A} - d_{tvm}) \cdot t} 
\end{align}

The correct fit value for $d_{rar}$ (which cannot be known in advance) equates the expected stream to the actual stream. Call this $d_{rarF}$. To compute it, we only care about the growth terms in eq. \ref{eq:Etvm_Atvm2}. Thus:
\begin{align}
g_{_E}-d_{tvm} - d_{rarF} &= g_{_A} - d_{tvm} \qquad 
\text{where: } 0 < g_A < g_E\\
g_{_E}-d_{rarF} &= g_{_A} \\
d_{rarF} &= g_{_E} - g_{_A}  > 0   
\end{align}

Therefore, because we are risk adjusting for $g_{_A}$ being lower than $g_{_E}$, $d_{rarF}$ is positive, as expected, and as accords with standard practice.

Now let us  consider a calamity, with an \textbf{expected} \emph{negative} stream of damages $D_E$ and an \textbf{actual} negative stream of damages $D_A$. The \textbf{loss streams} ($L)$ start at the same initial value $L<0$. These damages grow over time at the rates $g_{_E}$ and $g_{_A}$. And as before, a \textbf{time value of money} discount $d_{tvm}$ is applied to both loss streams, yielding $D_{E_{tvm}}$ and $D_{A_{tvm}}$, since losses at a future date are less concerning than losses in the present.
\begin{align}
D_E(t)&=L \cdot e^{(g_{_E}) \cdot t}   &  D_A(t)&=L \cdot e^{(g_{_A}) \cdot t} \\
D_{E_{tvm}}(t)&=L \cdot e^{(g_{_E} - d_{tvm}) \cdot t}   &  D_{A_{tvm}}(t)&=L \cdot e^{(g_{_A} - d_{tvm}) \cdot t} 
\end{align}
In the case of future losses, the risk is that these will be \emph{larger} than expected. We therefore seek to find a \textbf{risk adjustment rate} value of $d_{rar}$ to account for true losses of a larger magnitude than expected losses, yielding $D_{E_{rar}}$. As before, $d_{rar}$ is applied only to \emph{expected} losses, since $D_{A_{tvm}}(t)$ is the benchmark we seek to match:
\begin{align} \label{eq:Zero_gt_ActualLoss_gt_Expected}
\begin{split}
&D_{E_{rar}}(t) =L \cdot e^{(g_{_E} - d_{tvm} - d_{rar}) \cdot t} \qquad   D_{A_{rar}}(t) =  D_{A_{tvm}}(t)=L \cdot e^{(g_{_A} - d_{tvm}) \cdot t} \\ 
&\text{Where: } 0>D_{E}(t) > D_{A}(t) 
\end{split}
\end{align}

As in the positive income stream case, we are only interested in the growth terms of eq. \ref{eq:Zero_gt_ActualLoss_gt_Expected}. We are again interested in finding the  fit for $d_{rar}$, (\textbf{risk adjustment rate fit}) which is $d_{rarF}$.
\begin{align}
g_{_E}-d_{tvm} - d_{rarF} &= g_{_A} - d_{tvm} & \text{Where: } g_A < g_E < 0 \\
g_{_E}-d_{rarF} &= g_{_A} \\
d_{rarF} &= g_{_E} - g_{_A} < 0 
\end{align} 

Therefore, because the growth rate of \textbf{actual} damages, $g_{_A}$, exceeds those of \textbf{expected} damages $g_{_E}$, then $d_{rarF}$ will be a \emph{negative} quantity. Since this term is subtracted from the exponent, the effect is to make the total discount smaller, not larger, when risk is taken into account. Thus it appears that Stern \cite{Stern2007EconomicsofClimateChangeSternReview} and Dasgupta \cite{Dasgupta2004Discounting} are correct in contending that a lower discount rate should be used, and this supports Dasgupta's proof that discount rates can be negative \cite[p. 282]{Dasgupta2021Discounting}. We further note that severe damages can lower GDP and that lower GDP is a mechanism for reaching negative discount rates. This proof was first published as a pre-print \cite{Hanley2022SignOfRiskForFutureLosses}.

\subsubsection{Implementation of present value of future losses ($PV_{FL}$)}
In the context of climate/weather damages, the PV concept is significantly different from the usual evaluation of future profits that PV was created to model. Consequently, we will refer to present value of future losses ($\mathbf{PV_{FL}}$) to emphasise this. Here we will assume that the nominal maximum DICE discount rate of 5.1\% \cite{Nordhaus2018ChangesinDICE_Model} should be interpreted as the result of a risk multiple of a base discount rate. We assume that this base rate is $d_{M{T30yr}}$, which is the highest base discount rate we can justify. From these assumptions we generate a risk factor ($r_{DF}$) which is  used to create a corrected, DICE based, risk adjustment discount rate $d_{r_{DICE}}$. Because the sign of risk is negative, (i.e. damages get larger instead of smaller) rather than multiplying by the $r_{DF}$ factor, we must divide $d_{M{T30yr}}$ by $r_{DF}$.

\begin{flalign}
  \begin{aligned} \label{eq:PVInterestRates}  
  d_{M{T30yr}} &= 1.57\% \qquad d_{DICE} = 5.1\%  \qquad  d_{M{Fed}} = 1.427\%  \\
r_{_{DF}}&={\frac{d_{DICE}}{d_{M{T30yr}}}} = 3.2484 \\ 
d_{r_{DICE}} &= \frac{d_{M{T30yr}}}{r_{DF}} = 0.4833\% \\
\text{Where: } d_{M{T30yr}} &= \text{ Mean inflation  adjusted par real yield rate for 30 year bond} \\
d_{DICE} &= \text{ Nominal DICE  2016R  discount  rate } \\
 d_{M{Fed}}\, &= \text{ Mean inflation  adjusted Federal  Reserve discount rate.}\\ 
r_{_{DF}} &= \text{ Corrected DICE 2016R risk factor. Assumes that DICE } \\
& \quad \text{  discount rate (5.1\%) is a risk multiple of base discount rate. } \\
d_{r_{DICE}} &= \text{ Corrected DICE 2016R risk adjusted discount rate} 
\end{aligned}
\end{flalign} 

 In figure \ref{figPVDrateNumYears} the impact of $d_{DICE}$ versus corrected discounting $d_{rDICE}$ (Tab B) is seen on the discount rate scale. The $d_{DICE}$ range of 4\% to 5.1\% is $\approx$2 standard deviations (dark grey isosurface) above the 1.57\% $d_{M{T30yr}}$ base discount rate. 

\subsubsection{$\mathbf{PV_{FL}}$ weather damages isosurface graph for 2025 CE}

\begin{figure}[!ht]
\centering\includegraphics[width=5.0in]{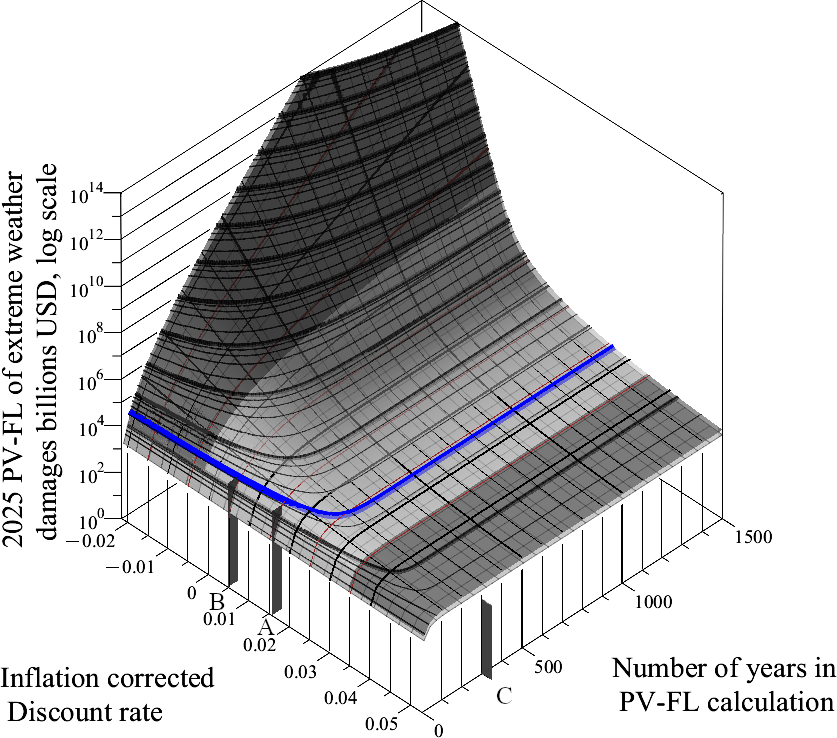} \\ 
\caption{\textbf{Isosurface graph:  $\pm$2$\sigma$ discount rate mapping of \textbf{$PV_{FL}$} for weather damages going forward from 2025. Represents total USA damages for all greenhouse gases together for the year 2025. } 
 The $\mathbf{x}$ axis scale varies the discount rate. The $\mathbf{y}$ axis scale is the number of years included in the $\mathbf{PV_{FL}}$ calculation. Losses are projected by the relevant function (\href{https://doi.org/10.48550/arXiv.2601.06085}{CMP:}  \S11.2) forward from 2025 divided by the discount rate per eq. \ref{eq:PVFLisosurface1}. 
Vertical \textbf{Tab A} at $\mathbf{d_{M{T30yr}}}$ of 1.57\% is the central rate from which $\sigma$ fluctuation is shown. The \textbf{Tab B} at $\mathbf{d_{r_{DICE}}}$ of 0.483\% discount shows the corrected DICE maximum and may be the most accurate net. The $\sigma$ derives from US 30 year T-bill real rate history (\href{https://doi.org/10.48550/arXiv.2601.06085}{CMP:}  \S3.4). Dark grey \textbf{Tab C} is the 300 year term of DICE. Light grey region centred at \textbf{Tab A} 1.57\% discount $\mathbf{d_{M{T30yr}}}$ is the centre of $\pm$$1\sigma$. Dark gray regions extend to $\pm$ $2\sigma$, and encompass the +5.1\% maximum DICE 2016R model rate (eq. \ref{eq:PVInterestRates}) and the -2.2\% rate. These outer $\pm$ 2$\sigma$ dark grey isosurface segments may be hard to believe, but given the potentially hundreds of years in which energy poverty (\S\ref{EnergyMoneyRelated}) or fiscal austerity and the paradox of thrift \cite{Corden2012GlobalImbalancesAndTheParadoxOfThrift,Vermann2012ParadoxThrift,Yiannis2011DeficitReductionTheAgeOfAusterityAndTheParadoxOfInsolvency,Das2019ParadoxOfAusterityMulti-CountryEvidence} could be created, such rates are not out of the question, and if we are to consider it possible that a 5.1\% real discount rate could exist, we should equally accept a -2.2\% discount. This latter value would be expected in an austere or energy poverty period, with a probable root cause of low energy availability. The impact of low energy availability is likely to be societal breakdown as future damages outpace capacity for repair and rebuilding.  
\textbf{Heavy blue contour line:} 2025's estimated GDP of \$24.7 trillion.The \textbf{minor Z log-scale contours} are 2, 4, 6, and 8 times the previous major contour.
}
\label{figPVDrateNumYears}
\end{figure}

The significance of this section is twofold. First, the 3 dimensional phase space for greenhouse gases cost is seen. This section also speaks to the practicality of pricing the effect of carbon and other GHGs into the economy. This applies equally to a carbon bond or a simple Pigouvian tax\footnote{A Pigouvian tax is assessed against private individuals or businesses for engaging in activities that create adverse side effects for society. Adverse side effects are those costs that are not included as a part of the product's market price. \cite{Kagan2023Pigouvian} }. Here we present graphical estimates of pricing carbon into the economy, though it is restricted to the United States. To this end the figure \ref{figPVDrateNumYears} isosurface graph of ${PV_{FL}}$ from weather damages shows the effects of discount rates and time span on total cost of carbon relative to 2025's GDP. These cover ${PV_{FL}}$ starting in 2025 CE over the $\pm$2$\sigma$ discount rates centred on the $\mathbf{d_{M{T30yr}}}$ 1.57\% discount rate. Previous climate economic analyses, such as Dietz' \cite{dietz2021tipping}, have used the high end 4\%-5.1\% discount rates, and low end span of years. 

\begin{flalign}
  \begin{aligned}  
\label{eq:PVFLisosurface1} 
PV_{FL}(fWD_m,d,k,n) &= \sum_{i=k}^{k+n} \frac{fWD_m(i) }{(1+d)^{i-k}} \cdot G_y \\ 
\text{Where: } fWD_m=& \text{ scenario function (\href{https://doi.org/10.48550/arXiv.2601.06085}{CMP:}  \S11.2)}  \\
 d=& \text{ discount rate } \\
 k=& \text{ starting year }\\
 n=& \text{ number of years to sum}\\
 G_y=& \text{ tonnes of GHG emitted in year }\textit{i}
\end{aligned}
\end{flalign}

 Figure \ref{figPVDrateNumYears} was generated using the above eq. \ref{eq:PVFLisosurface1} for 2025's $\mathbf{PV_{FL}}$ and varying the number of years and discount rate. This figure totals the damages costs with total quantities of GHGs. We start with a minimum number of years $\mathbf{PV_{FL}}$ period of 10 years, which corresponds to a $\mathbf{PV_{FL}}$ cost summation termination year of 2035 CE. The values along the contour from $\mathbf{C}$ show the 300 year cost summation, which is the approximate DICE time period \cite{dietz2021tipping,nordhaussztorc2016Rdice,Nordhaus2019TheUltimateChallengeForEconomics}. The maximum cost summation time period is 1500 years from 2025 (3525 CE). The scenario functions sum costs modified by remainder of all GHGs activity for each  year going forward. 
 
 The central light grey region of the isosurface is centred on $\mathbf{d_{M{T30yr}}}$ 1.57\% (\href{https://doi.org/10.48550/arXiv.2601.06085}{CMP:}  \S3.4.), and extends  standard deviation ($\pm1\sigma$) higher and lower. There are two darker grey regions next to the central 1$\sigma$, extending $\pm2\sigma$ on each side. Here, $\sigma$ is the standard deviation of the US 30 year T-bills used as the empirical basis for $d_{MT30yr}$. (\href{https://doi.org/10.48550/arXiv.2601.06085}{CMP:}  \S2.4.) 

 Figure \ref{figPVDrateNumYears} shows how wide a range of real world costs can result from the same physical world inputs. In the real world, economic conditions will vary. Real-dollar discount rates will not stay fixed. For most purposes, this means that a real-world path will be something like a random walk on this surface. The high cost regions correspond to economic conditions like energy poverty (\S\ref{EnergyMoneyRelated}) and/or austerity, with austerity's paradox of thrift \cite{Corden2012GlobalImbalancesAndTheParadoxOfThrift,Vermann2012ParadoxThrift,Yiannis2011DeficitReductionTheAgeOfAusterityAndTheParadoxOfInsolvency,Das2019ParadoxOfAusterityMulti-CountryEvidence}. 

\subsection{Willingness to pay (WTP) in the USA}
To address the criticism of Pindyck that IAMs are ad hoc, and the caveats of Manne which are discussed above in section \S\ref{sect:DICE_MERGE_WTP}, we took a bottom-up approach to the concept of willingness to pay, replacing Manne et al \cite{manne1995merge,manne2005merge} who used 2\% of 1995's GDP as the willingness to pay value which was adopted by DICE. This 2\% of GDP value was the environmental spending budget in the USA for 1995, and Manne expressed serious caveats in his paper. Dietz interprets Manne, graphing results with sigmoid curves stating a WTP high of 5.1\% to avoid +4°C, and just 2\% of GDP to avoid +2.5$\degree$C of warming \cite[p.44, fig 9]{dietz2021supplement}. (\href{https://doi.org/10.48550/arXiv.2601.06085}{CMP:}  \S8.8). 
  
\subsubsection{Our approach: WTP from historical  government crisis spending}

We base WTP on what the US government has allocated to deal with major crises as a realistic basis for estimates. For these we use the defence increase for WW2 (1941-1945)\cite{Tassava2008AmEconWW2}, Korean War (1950-1953)\cite[1950-1953]{FRED1947-2021FDEFX}, and Iraq/Afghan war (2003-2011)\cite[2003-2011]{FRED1947-2021FDEFX}. We also reference quantitative easing (QE) for the 2008 Global Financial Crisis \cite[2008-2011]{FREDqe2008}, QE for the COVID-19 pandemic \cite[2020-2021]{FREDqe2008}, and Federal spending for the 2020 pandemic \cite{FRED2020FGEXPND}. With the caveat that energy is foundation to real monetary valuation\cite{JARVIS2018EnergyReturns_TheLong-runGrowthOfGlobalIndustrialSociety,KEEN2019NoteOnEnergyInProduction}, we argue that spending to meet challenges shown in Table \ref{TableMonRespCrisis} more reasonably reflects limits to meet a crisis that is in the class of climate change. This economic support comprises spending, support for the financial sector, and bad debt retirement. These types of economic support operate together to achieve much greater total money provision.

QE's effect on the money supply depends on what is purchased. Central bank (CB) purchases of bonds from banks is a near net zero operation, providing a small profit. Banks purchasing bonds from the CB using funds from the CB discount window is a money supply expansion operation \cite{Jefferson2024SpeechFedDiscountWindow1913–2000,Ihrig2020FedsNewMonetaryPolicyTools,BGFRS2016FAQPrintingMoney}. CB purchase of junk assets from institutions expands (or, arguably maintains) the money supply when above fair value is paid for junk assets \cite{Delgado2023CentralBankAssetPurchases}. Thus, the CB can act as de facto insurer of last resort, as was done in response to the 2008 banking crisis to rescue the banking system from AIG's collapse \cite[pp. 14-19]{Hanley2012ReleaseOfTheKraken}. 

 \begin{table}[hbt!]
 \caption{WTP: Size of monetary response to crises in USA post WW2} 
\label{TableMonRespCrisis}
\begin{tabular}{lrrr}
                     & Event \% change &  Event     &  \\
Event                & from baseline   & \% of GDP &  \\
\hline
WW2                  & 2914.46\%       & 29.95\%   &  \\
Korean War           & 240.40\%        & 35.54\%   &  \\
Iraq/Afghan War      & 165.84\%        &  8.28\%    &  \\
QE banking crisis 2008-2011         & 267.19\%        & 9.98\%    &  \\
QE pandemic 2020              & 192.41\%        &  17.81\%   &  \\
Pandemic 2020 spending & 143.61\%        & 10.25\%   &  \\
\hline
Mean                 &                 & 18.64\%   &  \\
Median               &                 & 14.03\%   &  \\
Maximum              &                 & 35.54\%   &  \\
Minimum              &                 &  8.28\%    &  \\
$\sigma$                    &                 & 11.55\%   & 
\end{tabular}
\end{table}

Federal spending for climate change comprises tax rebates, tax deductions, contracts, and direct purchases of goods. Tax rebates and deductions are delayed by a year or more from the time of decision, which lowers the cost. Both rebate and deduction spending is amplified by the fraction of total money creation in the private sector that the program supports, typically a factor of 3-4 times, and also by increased monetary velocity. (However, modern money has quite low velocity \cite{FRED2024M1V,FRED2024M2V}.) Government contracts may last for 5 to 10 years or more and payments are usually for milestones through the contract's life. A company with a contract can immediately get a line of credit, without government expenditure until one or more years in the future when progress payments are made. Only direct purchases create immediate expenditure. 
  
The central bank can provide loan guarantees as directed, or de-facto retire bad debts of banks (QE). As we see in table \ref{TableMonRespCrisis}, QE is a powerful tool, and it requires no direct spending by government. 
 
The mean of table \ref{TableMonRespCrisis} is 3.65 times the 5.1\% maximum shown in Dietz, et al \cite[p. 44]{dietz2021supplement}. The table shows a range from 1.62X to 6.97X the Dietz' maximum. Dietz' like most of climate economics appears to use a descriptivist marginal assumption (See Stern 2007 \cite[pp.41-54]{Stern2007EconomicsofClimateChangeSternReview}).

\subsection{Social Cost of greenhouse gases is a multi-dimensional phase space for each year.}

\begin{figure}[!ht]
\centering\includegraphics[width=5.6in]{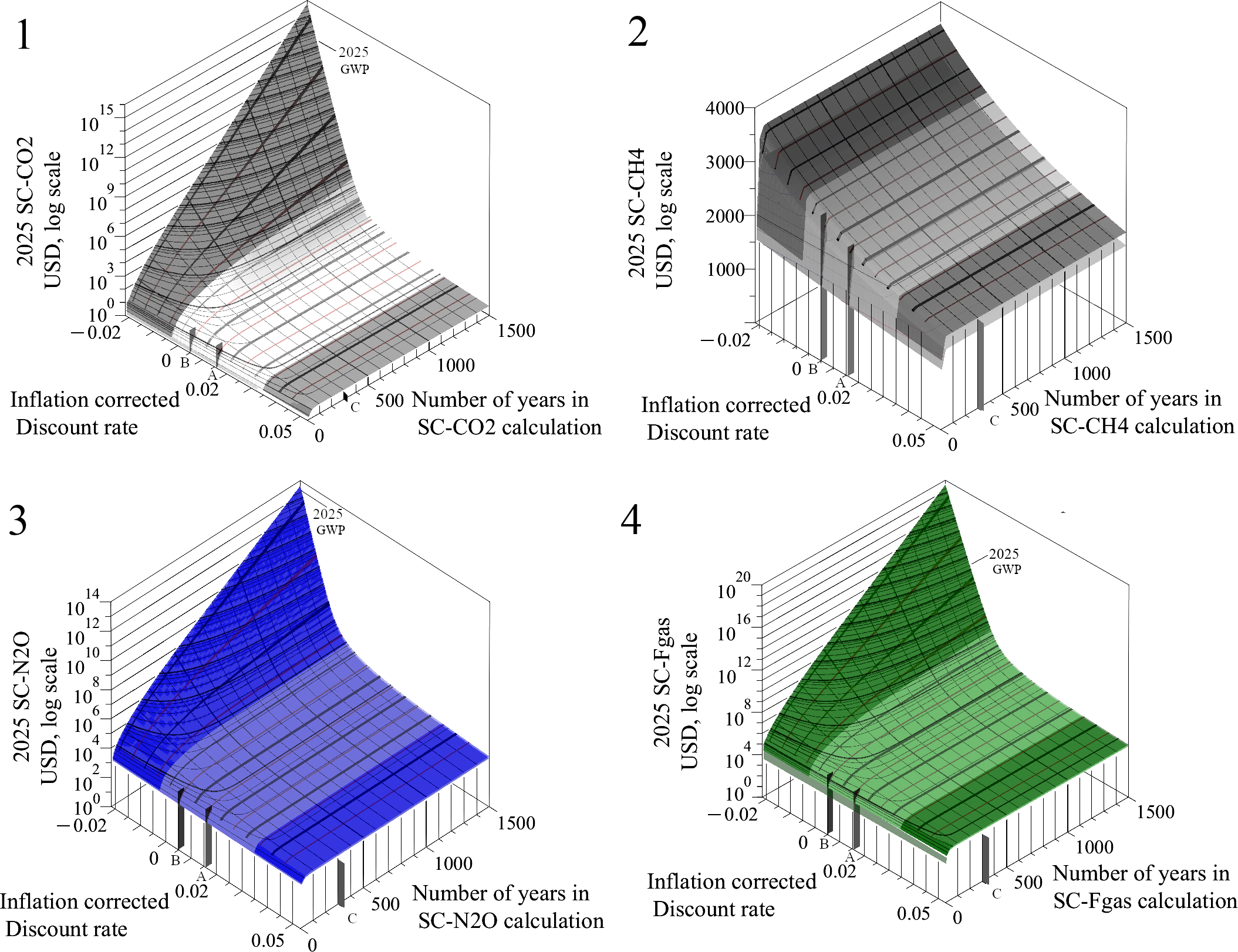}
\caption{\textbf{Social cost (SC) of emitting 1 tonne of each GHG in 2025.}  The \textbf{x} and \textbf{y} axes are, respectively, real-dollar discount rate and the number of years included in the computation. \textbf{Discount rate: -2$\sigma$ to +2$\sigma$ below and above Tab A. Baseline low scenario, and S-aerosol high CO$_2$ permafrost melt scenario---together provide the thickness of the isosurface.} From upper left, \textbf{Graph 1} is SC-CO$_2$, \textbf{Graph 2} is SC-CH$_4$, \textbf{Graph 3} is SC-N$_2$O, and \textbf{Graph 4} is SC-Fgas (halogenated gases, primarily flouro-chloro-carbons). At this scale the thickness of the isosurfaces is barely visible except for Graph 2 SC-CH4. The projected 2025 gross world product (GWP) is presented for reference where applicable. Discount scale marker tabs: \textbf{Tab A} is carbon bond mean discount rate $d_{MT30yr}$ = 1.57\%; \textbf{Tab B} is $d_{rDICE}$ = 0.483\% discount rate.  Range of discount rates is centred on \textbf{Tab A}'s $d_{MT30yr}$ discount rate, with a $\pm$ 2$\sigma$ range, where standard deviation is derived from analysis of 20 and 30 year treasury bill rates. (\href{https://doi.org/10.48550/arXiv.2601.06085}{CMP:}  \S3.4) This 2$\sigma$ range is chosen so that it encompasses the DICE model standard discount rate range of 5.1\%. It is unlikely to be realistic to presume a fixed discount rate, so the real world should be expected to take a wandering path relative to this isosurface---keeping in mind that each year has phase space graphs like this. Discount rates are in real dollars.  \textbf{Tab C} is 300 year nominal term of our carbon bond notes, and the approximate time span of the original DICE model. The \textbf{minor Z log-scale contours} are 2, 4, 6, and 8 times the previous major contour. Individual large versions for each GHG are provided in \href{https://doi.org/10.48550/arXiv.2601.06085}{CMP:} \S12} 
\label{Fig_SCC_isosurface_2Views}. 
\end{figure}

In this heat-conjecture model we compute an empirically founded cost per tonne of emitted GHGs (Fig. \ref{Fig_SCC_isosurface_2Views}). Thus, our "SCC" is a set of SC-GHG's, and represents a more sophisticated view. Each SC-GHG is at least a 3 dimensional phase space for each of: SC-CO$_2$, SC-CH$_4$, SC-N$_2$O, and SC-Fgas. This cost is subject to the limits of costs that are present in the TWD, and can be amended based on work on what costs are included in the TWD.

To compute the SC-GHG basis, the cost of a tonne of GHG is computed based on our scenarios and time span, per equations (\href{https://doi.org/10.48550/arXiv.2601.06085}{CMP:}  \S11.1-11.8). These results represent our best effort, validated by fitting existing climate data, excluding most potential tipping points and other surprises (\href{https://doi.org/10.48550/arXiv.2601.06085}{CMP:}  \S5). The surfaces in figure \ref{Fig_SCC_isosurface_2Views} do not include stochastic variance risk, nor all believed climate risks.  

Large, higher resolution individual isosurface graphs are provided in \href{https://doi.org/10.48550/arXiv.2601.06085}{CMP:} \S12. Summary GHG tables are provided in \href{https://doi.org/10.48550/arXiv.2601.06085}{CMP:} \S13, and full tables are in the supplementary depository. 

\section{Discussion}
\label{lb:Discussion}
A key takeaway of figure \ref{Fig_SCC_isosurface_2Views} is the same as that of the USA-centric figure \ref{figPVDrateNumYears}. (The latter is total real-dollar cost of all GHG emissions in 2025; the former is social cost of 1 tonne of GHG emitted in 2025.) The real-dollar discount rate of carbon bonds have a high impact on affordability of what we must do to deal with climate change. Allowing the real discount rate to be 0\% (no bond) or negative could be devastating. What the peak to the low side of the Tab B's 0\% discount means is breakdown of civilisation as we know it if it is maintained. It means society is unable to keep up with damages.

Consensus on the discounting controversy is intractable (\href{https://doi.org/10.48550/arXiv.2601.06085}{CMP:} \S2). Our analysis has to agree that the prescriptivist view is correct \cite[pp.41-54]{Stern2007EconomicsofClimateChangeSternReview}, which sets risk-free discounting at 0\%. Per our proof (\S \ref{Risk_proof}), adding risk makes the discount go below zero. However, looking through a bond-based discount rate lens, the prescriptivist versus descriptivist collision does not matter. We have cut the Gordian knot by use of a bond mechanism for discounting. It appears that discount rate achieved by the bond mechanism may be one of the largest factors, and it is certainly one we have control over. 

\subsection{Uncertainties include span of years and discount rate}

GHG emissions uncertainties are most visible in figure \ref{Fig_WD_Heat_Conjecture_graph}, which is a single slice through the 1.57\% discount rate. It is harder to see in graphs of social cost (fig. \ref{Fig_SCC_isosurface_2Views} \& \href{https://doi.org/10.48550/arXiv.2601.06085}{CMP:}  \S12). Figure \ref{Fig_SCC_isosurface_2Views} shows thickness of the lowest to highest projections. However, only in figure \ref{Fig_SCC_isosurface_2Views}-2 (SC-CH$_4$) is the thickness easily visible, because CH$_4$ impact is so limited. The uncertainty of GHG emissions is mostly overshadowed by two other factors: span of years and discount rate. 

The span of years over which to sum the residual effect of a GHG after its year of emission can have a large effect on the social cost for that year (fig. \ref{Fig_SCC_isosurface_2Views}), but this is dependent on the discount rate.  The largest potential uncertainty lies in the discount rate. We centred the discount rate scale on the mean inflation adjusted T-bill discount rate (fig. \ref{Fig_SCC_isosurface_2Views}[Tab A]) of 1.57\%. On either side of Tab A, shading shows 1$\sigma$ (light color) and 2$\sigma$ deviation (dark color) from the bond mean (\href{https://doi.org/10.48550/arXiv.2601.06085}{CMP:}  \S3.4-\S3.5). This most important factor, discount rate, humanity should largely be able to control. 

For OPTiMEM's heat-conjecture SC-GHG estimates, the higher the real interest rate paid by long term carbon bonds, the lower the impact on humanity. However as figure \ref{Fig_SCC_isosurface_2Views} makes clear, above the mean real dollar discount rate of 1.57\% at Tab A, there are diminishing returns on higher discount rates. On the positive side, this suggests bonds can be practical. On the other hand, in the negative discount region below Tab B's 0\% discounting, matters become intolerable very quickly. This should correspond to nation in energy poverty (\S\ref{EnergyMoneyRelated}) or austerity policy with potential paradox of thrift \cite{Corden2012GlobalImbalancesAndTheParadoxOfThrift,Vermann2012ParadoxThrift,Yiannis2011DeficitReductionTheAgeOfAusterityAndTheParadoxOfInsolvency,Das2019ParadoxOfAusterityMulti-CountryEvidence}. These can cause nations to be unable to keep up with damages, and create inflation above desirable rates.  

\subsection{There is not a single cost per tonne of emitted GHG}
Because OPTiMEM's heat-conjecture SC-GHG model is projecting from the current era using empirical realism, with 18 scenarios to model uncertainty, and that uncertainty is compounded by the choice of span of years interacting with real-dollar net discount rate, we cannot generate a single cost of carbon, and we must question the concept of a single social cost for any GHG. 

Increasing year-on-year SC-GHG values must occur because global heat content has an increasing trend, year-on-year, and heat provides the energy to drive damages. On top of this, cost can increase because of discount rate fluctuations and span of years. We submit that the true span of years may be very long indeed \cite{Archer2020UltimateCostofCarbon}. 

\subsection{The CO$_{2}$ equivalent concept is dangerously flawed}
Initially, this modelling effort attempted to use CO$_2$ equivalents within OPTiMEM. It became apparent that this was simply impossible because CO$_2$ has a unique chemistry. It dissolves into the ocean, which will then release CO$_2$ when the concentration declines. CO$_2$ is taken up, and released by rock. The behaviour of CO$_2$ over both short and long time scales is unique (fig. \ref{Fig-RemainderFractions} \cite{Archer2020UltimateCostofCarbon}. CH$_4$ has a half-life/\emph{e}fold conversion to CO$_2$. The other gases fit half-life/\emph{e}fold equations (\href{https://doi.org/10.48550/arXiv.2601.06085}{CMP:}  \S6.3). 

Given the longevity of the climate problem, the CO$_2$ equivalent metrics tend to overstate the climate impact of other GHGs, leading to unwarranted emphasis on GHGs like CH$_4$ \cite{Tans_2022UseAbuseOfC-14andC-13InAtmosphericCO2}. It is hard to imagine a gas less suited to being used as an equivalence metric for climate damages than CO$_2$. 

Consequently, we are strongly opposed to the CO$_2$ equivalent concept, only using it here as an inaccurate shorthand because it is used in other sources.  That said, figure \ref{Fig_SCC_isosurface_2Views} allows some reasonable basis of comparison. However, even this should be examined in a more sophisticated manner because of cost/benefit trade-offs, and the relative contributions to the total problem (\href{https://doi.org/10.48550/arXiv.2601.06085}{CMP:}  fig. 37). At this time, and for the next century to come, CO$_2$ is the problem to solve. We can keep an eye on other GHGs. 

For instance, N$_2$O has a high social cost per tonne. It certainly makes sense to prioritise industrial abatement of N$_2$O immediately because it is straightforward and does buy us a fair amount in the far future. But given how critical nitrogen is for agriculture, agricultural abatement should be done carefully. There is plenty of time to study N$_2$O while accomplishing very significant abatement from industry, and figuring out how to approach it, perhaps in tandem with the development of robotic farming multi-cropping. Modelling of  Anderson's ocean halogen ozone effect should be done for N$_2$O \cite{Anderson2017ChlorineBromineCatalysis,Anderson2018CouplingCatalysis}. 

\subsection{What if no carbon bonds are issued?}
Looking at figure \ref{Fig_SCC_isosurface_2Views} one can ask the question, "What if no carbon bonds are issued?" Viewed in this light, the effect of the x-axis discount rate of 0\% can be interpreted as showing us what the range of outcomes can be. However, treasury bills are regularly issued and the proceeds used to pay for the needs of government, and other central banks do the same thing. Thus, the investments of governments around the world to deal with climate change could be modelled as equivalent to these carbon bonds. However, our carbon bond term is 10 times the maximum term of a US treasury bill, which is supported by experience of the Bank of England (\href{https://doi.org/10.48550/arXiv.2601.06085}{CMP: \S3}). And, the spending made possible by the bond must be effective at energy replacement with high EROEI sources. 

Viewed in this way, figures \ref{figPVDrateNumYears} and \ref{Fig_SCC_isosurface_2Views} can be seen as a guide to how much governments should be spending on truly effective energy source changeover. The observations about energy poverty (\S\ref{EnergyMoneyRelated}) and austerity, the latter inducing the paradox of thrift \cite{Corden2012GlobalImbalancesAndTheParadoxOfThrift,Vermann2012ParadoxThrift,Yiannis2011DeficitReductionTheAgeOfAusterityAndTheParadoxOfInsolvency,Das2019ParadoxOfAusterityMulti-CountryEvidence}, both potentially steering nations onto the peak below 0\% remain true. 

\section{Concluding remarks}

It is of paramount importance that climate economics become congruent with  climate science. Both fields must work hard to understand each other, and failure to do so should not be acceptable. Climate science does not  comprehend the approaches, metrics, and goals of climate economics, and this must change. Some within the field of climate economics, understand that there are problems, but are unsure of what to do about it. The field of climate economics does not understand climate science, for instance continuing to apply temperature as scalar increases when there is not the straightforward relationship between temperature and damages assumed, and this fault must not continue. Our modelling effort is intended to begin a bridge between these almost entirely disjoint fields. We hope that this bridge begins the process of education that is urgently needed. 

Climate economics needs to understand these fundamental curves of: carbon combustion $\Rightarrow$ GHG increases $\Rightarrow$ $\Delta$T (temperature change) $\Rightarrow$ $\Delta$Q (ocean heat change) $\Rightarrow$ total OHC (\href{https://doi.org/10.48550/arXiv.2601.06085}{CMP:}  \S6.5 \& fig.'s 45-46A).  Each curve lags the previous by decades or centuries. This begins the process of comprehending how climate damages might appear. These curves also show what should happen after net-zero occurs --- the release of tens of zettajoules of ocean heat as heat flow goes negative --- which should cause massive storms (\href{https://doi.org/10.48550/arXiv.2601.06085}{CMP:}  fig.'s 15 \& 45). Understanding fundamentals of weather dynamics, such as the ratio of energy between precipitation and wind (\href{https://doi.org/10.48550/arXiv.2601.06085}{CMP:}  fig. 41), that is weighted by orders of magnitude towards rainfall, helps to develop an intuitive grasp of increased evaporation. 

OPTiMEM provides a foundational alternative method for computing social cost of CO$_2$, CH$_4$, N$_2$O and halogenated hydrocarbon gases. Thus it can be added to the set of methods for such use. Our method for estimating risk of bad weather years at any frequency makes this model valuable for actuaries and engineers as well. On balance, the OPTiMEM heat-conjecture SC-GHG model is probably more likely to project low rather than high (\href{https://doi.org/10.48550/arXiv.2601.06085}{CMP:}  \S5, \S9). This is consistent with the inherent conservatism of science \cite{Herrando-Perez2019StatisticalLanguageBacksConservatisminClimateAssessments,BRYSSE2013ClimateChangePredictionErringOnTheSideOfLeastDrama,Hansen2023WarmingInPipeline}. Being more likely to project lower than higher should also follow from the fact that our model only accounts for direct damages (\href{https://doi.org/10.48550/arXiv.2601.06085}{CMP:}  \S4.4), and like most models will be refined \cite{Hanley2026ReviewOfNOAAsBillionDollarDisastersInlightOfritiquingRogerPielkeJrsCriticalRemarks}.

\subsection{Energy is the foundation of monetary valuation}
\label{EnergyMoneyRelated}
The policy of climate change should consider that the value of money is largely due to energy. Of course, inflation arises from multiple sources, from oil price \cite{HA2023UnderstandingGlobalDriversOfInflation_HowImportantAreOilPrices} which generally supports the energy thesis, to politics \cite{WHITEHEAD1979THEPOLITICALCAUSESOFINFLATION}. However, what happens when inflation is removed? Jarvis shows a strong relation between energy and the economy \cite{JARVIS2018EnergyReturns_TheLong-runGrowthOfGlobalIndustrialSociety}.  A strong relationship of energy to money makes intuitive sense. In physics work is defined as force $\times$ distance ($W=f \times d$), which is analogous to human and machine labour. Jarvis's finding supported revising the Cobb-Douglas Production Function (CDPF) to an energy-based CDPF (EBCDPF) \cite[pp. 44-45, eq. 1.22]{KEEN2019NoteOnEnergyInProduction}. This EBCDPF can be summarised: Energy is used by capital and labour to accomplish work leveraged by harnessing energy to productive use.  The end product is GDP, which creates prosperity for the nation. This is aliged with the thermodynamic relation between monetary value, energy consumption, and the net exergy efficiency of energy \cite{hanley2026newmonetarymetricthermodynamic}.

The implication of these findings about energy is that the idea that we can cut energy use by 10\%, 50\% or 90\% (all figures mentioned by advocates) and simplify our civilisation while maintaining a viable economy is not going to work, except to the degree greater improvements in efficiency occur. This finding means that energy poverty $\Rightarrow$ monetary/economic poverty. There is little reason not to pursue a degree of frugality---however, cue the paradox of thrift \cite{Corden2012GlobalImbalancesAndTheParadoxOfThrift,Vermann2012ParadoxThrift,Yiannis2011DeficitReductionTheAgeOfAusterityAndTheParadoxOfInsolvency,Das2019ParadoxOfAusterityMulti-CountryEvidence}. 

These observations regarding energy have profound implications for addressing climate and maintaining a civilization as we know it. It means that EROEI is the most crucial metric, ceterus paribus \cite{JARVIS2018EnergyReturns_TheLong-runGrowthOfGlobalIndustrialSociety}. 

\subsection{Creating a real-dollar effective discount rate could be a powerful tool}
Our carbon bond creates a robust real-world mechanism for climate discounting. Capacity to pay real-dollar returns depends on high EROEI, and the efficiency of energy use (exergy). This has made visible just how much effect such a real-world discount has on our ability to deal with climate change (fig. \ref{Fig_SCC_isosurface_2Views}). Of course this only works if the money raised to fight climate change is used optimally---spending must actually mitigate effects on future generations. The largest part of that mitigation is bequeathing robust energy generation capacity sufficient to allow human adaptation to the severe effects of climate change that will appear.  

Our modelling shows that financial tools, such as our carbon bond, should be of great importance for climate change. Requiring emitters to purchase carbon bond offsets may act as a type of carbon tax. Such a Pigouvian implementation for carbon bonds would inject accounting for future costs, with the goal to motivate free market changes.  Using carbon bonds in this way is reasonable when the price is well within capacity to pay, timelines are long, and markets are stable. However, as climate change accelerates, timelines become short and market stability may be doubtful. In an energy crunch, capacity to pay may become problematic, and perverse incentives to spike energy prices appear \cite{moncarz2006TheRiseAndCollapseOfEnron}. The world is in such a short timeline period---just until 2025 CE to achieve benchmarks \cite[SPM-21]{IPCC_AR6_WG3_Report} (or more realistically, 2035-2045, accepting significant catastrophe).

Policy makers should understand that when climate damages become large and energy becomes expensive, this drags real-dollar GDP down and negative net discount rates should result \cite{Dasgupta2021Discounting}, drastically raising the SC-GHG estimates (fig. \ref{Fig_SCC_isosurface_2Views}).  Similarly, attempting to price carbon directly into the economy is itself a significant drag on real GDP, although spending equal to that cost drag may be sufficient counterweight---if spent on effective replacement of fossil fuel. 

Practising austerity or energy poverty can increase, or create a downturn. Austerity can do so by the paradox of thrift \cite{Corden2012GlobalImbalancesAndTheParadoxOfThrift,Vermann2012ParadoxThrift,Yiannis2011DeficitReductionTheAgeOfAusterityAndTheParadoxOfInsolvency,Das2019ParadoxOfAusterityMulti-CountryEvidence}. Energy poverty does so by taking the foundation out of GDP, as production is based on energy (\href{https://doi.org/10.48550/arXiv.2601.06085}{CMP:}  \S1.5.7 \& \S3.4). Either practice should force net negative carbon bond rates, raising SC-GHGs (fig. \ref{Fig_SCC_isosurface_2Views}), in a self-perpetuating feedback cycle. That cycle, in turn, feeds back into society, enhancing socio-political factors that can militate against reduction efforts \cite{millward-hopkins_2022WhyImpactsLessReduceEmissions}. To prevent the physics based downturn cycle requires increasing energy capacity in step with provision of sufficient money to at least keep up with damages and ensure populations are employed and able to maintain living standards. This provides basis for the real economy of goods and services to grow \cite{KEEN2019NoteOnEnergyInProduction,JARVIS2018EnergyReturns_TheLong-runGrowthOfGlobalIndustrialSociety}. 

\subsection{Carbon bonds need to be used for low-cost high EROEI energy}

We propose that carbon bonds fund replacement of fossil fuels with the highest EROEI forms of energy production, which is clearly possible using nuclear power technology \cite{Lang2017NuclearPowerGlobalBenefitsForgone,LOVERING2023MicroreactorsOffgrid}.  With  a viable high EROEI foundation of cheap, plentiful energy, a 100\% synthetic hydrocarbon cycle may be supportable (\href{https://doi.org/10.48550/arXiv.2601.06085}{CMP:}  \S6.1.2). Electrification should go forward, but the energy density of synthetic hydrocarbons is difficult to match, and difficult to replace in many industrial situations. 

Policy makers could start tracking the amount of money spent on climate mitigation, the net real-dollar discount rate on the bonds sold to pay for it, plus the EROEI and total supply chain effectiveness of the spending at mitigating GHG. This can be compared to the amount of bonds that should be allocated each year to pay for the cost of emissions for each GHG each year. The energy capacity for support of GDP should also be tracked. Together this would provide a metric for measurement of effectiveness of climate mitigations. 

\subsection{Concerns about long-term security implicit in some low-energy proposals}
It is concerning that low-energy futures currently proposed  
\cite{Moriarty2020FeasibilityOf100pctGlobalRenewableEnergySystem,WEIBACH2013EROIandEPBTofElectricityGeneratingPlants}, will seriously weaken the nations that embrace it, leaving those that do not take the low-energy path to dominate the future---moral suasion does not dominate realpolitik. This is particularly true for the United States in context of the erosion of the US dollar's near monopoly on settlement of international trade \cite{Liu_Papa_2022CanBRICS_DedollarizeGFS, Papa2023BRICSconvergenceIndex}. Consequently, zealous avoidance of expensive, low EROEI energy to add to this burden is appropriate. 

\subsection{Will there be enough low-cost fossil fuel energy available for the massive energy generation build required?}
The process of estimating the limits of fossil fuel production for our baseline scenario highlighted the real possibility that the USA, and by extension, the globe, may not have enough low cost energy  \cite{Tverberg2021FossilFuelProblem} to build the high EROEI generation facilities required while maintaining living standards. Already, decreasing EROEI for fossil fuels is putting limits on utilisation \cite[figs. 5, 6]{Perissi2021TheRoleOfEROEIinComplexAdaptiveSystems}, and the globe may be close enough to this limit to put hard stops on fossil fuel \cite{Brockway2019EstimationOfGlobalFinalStageEROEIForFossilFuelsWithRenewableEnergy}. As part of this, we have added the burden of low EROEI energy development to the straightforward energy demands of the global economy, when the world needs replacement with high EROEI. 

The urgency of enacting effective policy is considerably greater than is generally believed. We are in basic agreement with Yang, et al \cite{Yang2021SolelyEconomicMitigationStrategySuggestsUpwardRevisionOfNationallyDeterminedContributions} regarding the probable requirements of mitigation, with the caveat that it will likely be higher, and necessary materials may not be available \cite{Michaux2024EstimationQuantityMetalsPhaseOutFossilFuels}. Nations need to become realistic about replacement of fossil fuels as rapidly as possible---to date, international responses are as realistic as the village of Potemkin. Adding to the quote below, one can note that providing support for the "less shining side", unwitting or no, can be a lucrative academic career. 

\begin{quote}
"The states coordination coin has two sides: talk, meetings, declarations, promises on the one side, and reneging, cheating and opportunism with guile on the other less shining side." - Lane 2015 \cite{Lane2015GHGEmissions_PotemkinVillagesAndGlobalResilience} 
\end{quote}

\subsection{Reasons for optimism}
There are reasons for optimism. Most critically, the \emph{capacity} of the US government to spend to meet this challenge, and by extension the \emph{capacity} of developed nations, appears to be quite sufficient, should the political will appear before low energy economic conditions appear---signalled by high cost of energy. Second, as long as fossil fuel cost remains within reasonable range costs should be affordable to build the nuclear conversion that is required \cite{LOVERING2016CostOfNuclearReactors}. A successful transition will require a wartime level of spending and removal of bureaucratic obstacles. The world is going to become a more difficult place to live in. We think, however, that if we do the right things, civilisation can survive and thrive. 
\\ \\

\backmatter

\bmhead{Supplementary information}
This manuscript has a \href{https://doi.org/10.48550/arXiv.2601.06085}{companion article (CMP:)} that is available on Arxiv. Scenario outputs available on Figshare \url{https://doi.org/110.6084/m9.figshare.31072156} Code and datasets available by request. 

\bmhead{Acknowledgements}
We wish to acknowledge Darrel Duffie for suggesting use of a bond to implement discounting, Tim Garrett, Martin Mlynczak, Philip Klotzbach, and James W. B. Rae, for datasets, weather references and guidance, David Archer for carbon model guidance, Gail Tverberg for discussion of actuarial limits, J. Paul Kelleher and Partha Dasgupta on PRTP, and to our survey responders for their participation.

\section*{Declarations}

\begin{itemize}
\item Funding {No external funding was received for this work.}
\item Competing interests {The authors assert no competing interests.}
\item Ethics approval and consent to participate {Not applicable}
\item Consent for publication {Not applicable}
\item Data availability {Available in the figshare.}
\item Materials availability {Not applicable}
\item Code availability (Availabley in the fig share.) 
\item Author contribution {Brian P. Hanley developed this approach inspired by work on a previous manuscript and was lead author. Pieter Tans coauthored and provided invaluable guidance, feedback and editing for the climate model and manuscript. Edward Schuur contributed invaluable guidance, feedback and editing of the permafrost melting section of the heat-conjecture SC-GHG model. Geoffrey Gardiner provided a foundation for the 300 year bond concept operating in the real world. Steve Keen contributed the mathematical proof of the sign of risk, and critique. Adam Smith contributed to the overall text, was the guide for use of the weather damages dataset as well as working on improving the NOAA dataset for use in this manuscript.}
\end{itemize}

\noindent


\bibliography{references2,Refs-+4C}

\end{document}